\documentclass[preprint2]{aastex}
\usepackage{float,graphicx,amsmath,multirow}
\usepackage{color}
\usepackage[final,addedmarkup=bf]{changes}

\title{CSO and CARMA Observations of L1157.  II.  Chemical Complexity in the Shocked Outflow}
\author{Andrew M. Burkhardt\altaffilmark{1}, Niklaus M. Dollhopf\,\altaffilmark{1}, Joanna F. Corby\altaffilmark{1}, P. Brandon Carroll\altaffilmark{2}, Christopher N. Shingledecker\altaffilmark{3}, Ryan A. Loomis\altaffilmark{4}, Shawn Thomas Booth\altaffilmark{1}, Geoffrey A. Blake\altaffilmark{2,5},  Eric Herbst\altaffilmark{1,3}, Anthony J. Remijan\altaffilmark{6}, Brett A. McGuire\altaffilmark{6,\dagger,*}}
\altaffiltext{1}{Department of Astronomy, University of Virginia, Charlottesville, VA 22904}
\altaffiltext{2}{Division of Chemistry and Chemical Engineering, California Institute of Technology, Pasadena, CA 91125}
\altaffiltext{3}{Department of Chemistry, University of Virginia, Charlottesville, VA 22904}
\altaffiltext{4}{Department of Astronomy, Harvard University, Cambridge, MA 02138}
\altaffiltext{5}{Division of Geological and Planetary Sciences, California Institute of Technology, Pasadena, CA 91125}
\altaffiltext{6}{National Radio Astronomy Observatory, Charlottesville, VA 22903}
\altaffiltext{$\dagger$}{BAM is a Jansky Fellow of the National Radio Astronomy Observatory}
\altaffiltext{*}{To whom correspondence should be addressed: bmcguire@nrao.edu}


\begin{document}
\begin{abstract}
L1157, a molecular dark cloud with an embedded Class 0 protostar possessing a bipolar outflow, is an excellent source for studying shock chemistry, including grain-surface chemistry prior to shocks, and post-shock, gas-phase processing. The L1157-B1 and B2 positions experienced shocks at an estimated $\sim$2000 and 4000 years ago, respectively. Prior to these shock events, temperatures were too low for most complex organic molecules to undergo thermal desorption. Thus, the shocks should have liberated these molecules from the ice grain-surfaces \textit{en masse}, evidenced by prior observations of SiO and multiple grain mantle species commonly associated with shocks. Grain species, such as OCS, CH$_3$OH, and HNCO, all peak at different positions \added{relative }to species that are preferably formed in higher velocity shocks or repeatedly-shocked material, such as SiO and HCN. Here, we present high spatial resolution ($\sim$3\arcsec) maps of CH$_3$OH, HNCO, HCN, and HCO$^+$ in the southern portion of the outflow containing B1 and B2, as observed with CARMA. The HNCO maps are the first interferometric observations of this species in L1157. The maps show distinct differences in the chemistry within the various shocked regions in L1157B. 
This is further supported through constraints of the molecular abundances using the non-LTE code \textsc{radex} \citep{VanderTak2007}. We find the east/west chemical differentiation in C2 may be explained by the contrast of the shock's interaction with either cold, pristine material or warm, previously-shocked gas, as seen in enhanced HCN abundances. In addition, the enhancement of the HNCO abundance toward the the older shock, B2, suggests the importance of high-temperature O-chemistry in shocked regions.
\end{abstract}

\section{Introduction}
\label{intro}
Shocks and turbulence are prevalent in molecular gas, from dense gas undergoing star formation to diffuse material filling the thin disk of the Galaxy. Thus, it is crucial to understand the underlying chemistry within shocked regions. For many complex molecules, gas-phase formation routes are inefficient under interstellar conditions, and are unable to reproduce observed abundances. Theoretical work instead predicts efficient chemical reactions on the ice mantles of dust grains \citep{HerbstReview,Brown88}, and laboratory work confirms that grain-surface chemistry efficiently produces diverse molecular species \citep{Palumbo15, Garozzo10}. Low-velocity C-shocks, common in dense molecular clouds, sublimate the ice from the surfaces of grains without destroying most molecular bonds \citep{RequenaTorres2006}. As a result, gas-phase observations of recently-shocked clouds can provide substantial information on the condensed-phase chemistry that occurred prior to the passage of the shock. Reactions on the surface of grains are now believed to be responsible for the formation of most complex organic species \citep{Garrod2013}, and studies of shocked molecular clouds are essential for understanding the processing of molecular material and the formation of biologically-relevant molecules. Given the difficulties of unambiguously identifying molecules trapped in ice grains, shock-chemistry studies are among the most promising cases for understanding ice grain chemistry.

The nearby ($\sim$250 pc) dark cloud, L1157, provides an excellent opportunity to study shock chemistry. L1157 contains a Class 0 protostar, L1157-mm, with a bipolar outflow that extends $\sim$2\arcmin, or $3\times10^4$ AU, from the central infrared source \citep{Looney2007}. The blueshifted and redshifted outflows are referred to as L1157B and L1157R, respectively. As the jet precesses, the periodic ejection events produce bow shocks at multiple positions in the outflow \citep{Gueth1996}. The bow shocks, in turn, produce regions of minimal protostellar heating of the dust, and significant shock-induced non-thermal desorption of grain mantles \citep{Fontani2014}. 

L1157-B1 and B2 are prominent, recently-shocked regions in the southern, blue-shifted lobe of the outflow. B1 is warmer and more recently shocked ($T_{\text{kin}}\sim$~80-100 K, t~$\sim$~2000 yr) than B2 ($T_{\text{kin}}\sim$~20-60 K, t~$\sim$~4000 yr) \citep{Gueth1996, Tafalla1995, Lefloch2012}. The progenitor gas for B1 and B2 should have had a cold history within the dark cloud (at T~\textless~20 K for $\sim$1~Myr). The unshocked portions of the outflow are also cool (at T$_{\text{kin}}$ $\sim$15 - 30 K) \citep{Bachiller2001}. As a result, observations of B1 and B2 primarily probe the chemistry of cold ($\lesssim$20 K) ice-grain mantles, and subsequent evolution in the warmer gas-phase. 

B1 and B2 are considered chemically-active outflows \citep{Bachiller2001}, with the detection of numerous complex species including HCOOCH$_3$, HCOOH, CH$_3$CHO, NH$_2$CHO, C$_2$H$_5$OH, and CH$_3$CN \citep[and refs. therein]{Arce2008,Yamaguchi2011}. Abundances of shock tracers including SiO, CH$_3$OH, and HNCO are particularly enhanced in B1 and B2 as compared to unshocked regions of the outflow. This includes an SiO abundance enhancement of $10^4$ compared to unshocked material in the region, which high spatial resolution ($\sim$2.5\arcsec) images indicate that the distribution of SiO closely traces the positions of B1 and B2 \citep{Gueth1998}. Furthermore, L1157-B2 has been found to contain the highest abundance of HNCO yet observed in any source in the Galaxy, including an enhanced abundance of 4-11 times compared to hot cores and dense cores in the Galactic Center  \citep{Rodriguez2010, Mendoza2014}. 

Interferometric images of multiple molecules have demonstrated the clumpy structure of the southern outflow.  Previous work by \citet{Benedettini2007} at $\lambda$=3 mm mapped transitions of CS, CH$_3$OH, HCN, OCS, and $^{34}$SO at $\theta\sim$ 3-6\arcsec\, spatial resolution with the Plateau deBure Interferometer (PdBI). In this work, we present images of 3 mm transitions of HCO$^+$, HCN, CH$_3$OH, and HNCO observed by the Combined Array for Research in Millimeter-Wave Astronomy (CARMA).  While two of the species have been imaged with comparable resolution, this work produces a higher spatial resolution map of HCO$^+$ and the first interferometric map of HNCO in L1157. 

Furthermore, we build upon the work of McGuire et al. (2015; hereafter Paper I) to estimate column densities and abundances of three of the species (CH$_3$OH, HCN, and HNCO) in the B1 and B2 clumps.  

\section{Observations}
\label{obs}
We observed the southern outflow with the CARMA 15-element array at $\lambda$~=~3 mm, targeting the $J=1-0$ transitions of HCN and HCO$^{+}$, four $J=2_K-1_K$ transitions of CH$_3$OH, and the $J=4-3$ transition of HNCO, as described in Table \ref{tab:carma_transitions}. A total of 89.3 hours of observations were conducted: 46.7 hours in C-configuration during August 2012, 23.3 hours in D configuration between October and November 2012, and 19.3 hours in E-configuration during May 2013. All observations were toward a single pointing position at $\alpha$(J2000) = 20$^{\mbox{h}}$39$^{\mbox{m}}$07$^{\mbox{s}}$.7, $\delta$(J2000) = 68$^{\circ}$01\arcmin11\farcs5. The 62 MHz bandwidth, 3-bit mode of the CARMA correlator was used, providing fourteen non-contiguous windows with 255 channels each, and resulting in 243~kHz ($\sim$0.7~km~s$^{-1}$) channel resolution.

The data were reduced using the {\sc miriad} package \citep{Sault1995} using standard techniques of bandpass, absolute flux, and complex gain calibration.  MWC349 and Neptune were used as primary flux calibrators; the bandpass calibrators were 1635+381, 2232+117, 0102+584, 1743-038, 2015+372, and 3C84; and the gain calibrator was 1927+739.  Images were generated using the {\sc clean} algorithm in the {\sc miriad} package with a Briggs weighting scheme that tended toward natural weighting, resulting in a synthesized beam of 3\farcs4~$\times$~3\farcs2 at the center frequency for the majority of the observed transitions. Image analysis was performed in {\sc casa}. The targeted transitions and synthesized beam values are shown in Table \ref{tab:carma_transitions}. The synthesized beam of the CH$_3$OH maps are about a factor of two larger than the other observed transitions because the CH$_3$OH lines were not observed in the most extended configuration used (C-array). While primary beam corrections were not applied, preliminary analysis implied that the ultimate impact of this correction would be minimal, especially for the regions where we perform more in-depth analysis. It should be noted that the sensitivity of the 10-m antennae will be more significantly reduced at the edges of the primary beam in this single pointing.

\begin{deluxetable}{c c c c c c}
\tablewidth{0pt}
\tablecaption{Targeted CARMA Transitions}
\tablecolumns{5}
\tablehead{\colhead{Molecule} & \colhead{Transition} & \colhead{$\nu$} & $E_u$ & \colhead{Beam} & \colhead{RMS ($\sigma$)} \\
& & (GHz) & (K) & (arcsec$^2$)& (mJy beam$^{-1}$)} 
\startdata
\multirow{4}{*}{\shortstack{CH$_{3}$OH \\ (methanol)}} 
    & $2_{-1,2}-1_{-1,1}$ & 96.73936(5) & 12.9 & \multirow{4}{*}{6\farcs03 $\times$ 5\farcs53} & \multirow{4}{*}{8.5} \\
	& $2_{0,2}-1_{0,1}++$ & 96.74138(5) & 6.9\\
	& $2_{0,2}-1_{0,1}$   & 96.74455(5) & 20.1\\
	& $2_{1,1}-1_{1,0}$   & 96.75551(5) & 28.0\\
\hline\noalign{\smallskip}
\multirow{3}{*}{\shortstack{HCN \\ (hydrogen cyanide)}}
    & $J=1-0, F=1-1$      & 88.63042(2) & 4.25  &\multirow{3}{*}{3\farcs45 $\times$ 3\farcs27} & \multirow{3}{*}{4.8} \\
	& $J=1-0, F=2-1$      & 88.63185(3) & 4.25\\
	& $J=1-0, F=0-1$      & 88.63394(3) & 4.25\\ \hline\noalign{\smallskip}  
\begin{tabular}{@{}c@{}}HCO$^{+}$ \\ (formylium)\end{tabular}
    & $J=1-0$             & 89.18853(4) & 4.28 & 3\farcs39 $\times$ 3\farcs23 & 5.1 \\ \hline\noalign{\smallskip}
\begin{tabular}{@{}c@{}}HNCO \\ (isocyanic acid)\end{tabular} 
    & $J=4_{0,4}-3_{0,3}$ & 87.92524(8) & 10.6 & 3\farcs50 $\times$ 3\farcs28 & 7.0
\enddata
\tablecomments{The resolved quantum numbers and frequencies of each transition are listed. Also included are the synthesized beam sizes and approximate rms levels for each window, where no spectral binning was performed. Transitions and parameters accessible at \url{www.splatalogue.net} \citep{Remijan2007}. Catalogued at CDMS \citep{Muller2005}. and JPL \citep{Pickett1998} }
\label{tab:carma_transitions}
\end{deluxetable}

\section{Results}
\label{results}

\subsection{Structure}
\label{results:structure}

\subsubsection{Summary of L1157B Structure}

\begin{figure*}
\includegraphics[width=\textwidth]{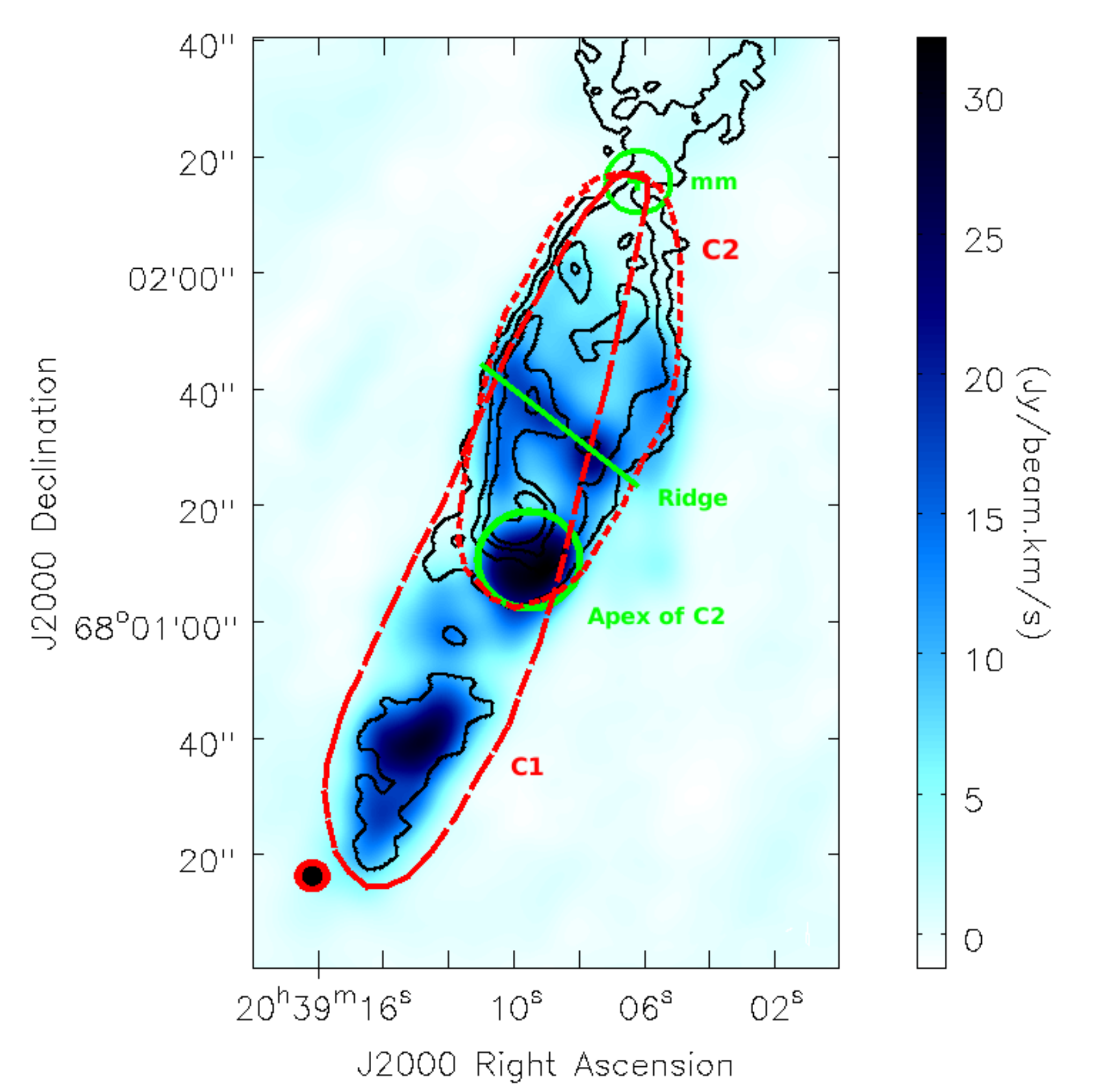} 
\caption{Integrated CH$_3$OH ($2_K-1_K$) emission raster, -17.8 to $+$14.0 km s$^{-1}$, with integrated emission of HCO$^+$ (1-0) contours, -20.1 to $+$11.9 km s$^{-1}$, overlayed in black, both observed with CARMA. The contour levels for HCO$^+$ are 0.44, 0.88, 1.32, 1.76, 2.2 Jy beam$^{-1}$ km s$^{-1}$. The synthesized beam sizes for CH$_3$OH and HCO$^+$ are shown in red and black, respectively, in the lower left. The two cavities in L1157B are shown and labeled in red, with the ridge and apex of C2 shown and labeled in green.}
\label{fig:structure}
\end{figure*}

To orient the reader to the structure of the southern, blueshifted outflow, L1157B, we highlight the positions of the shocked clumps with respect to the two cavities, C1 and C2, originally observed in CO \citep{Gueth1997}. Observations of SiO, a prominent shock tracer, reveal large abundances along the structures of the cavities. Figure \ref{fig:structure} shows line emission by CH$_3$OH and HCO$^+$ observed by CARMA with the position of the cavities highlighted. The southern-most cavity, C1, holds the shocked B2 region, while the B1 region is observed at the apex of the C2 (northern) cavity. The bowshock nature of the outflow is apparent from the U-shaped structure of C2 observed with CH$_3$OH, other shock tracers, and mid-IR emission \citep{Takami2010}

The B0 region follows the shape of the C1 cavity north of B1 into the L1157-mm source; however, B0 is not correlated with a shock front, and much of it remains cool \citep[$T_{\text{kin}}\sim$15-30 K;][]{Bachiller2001}. The aforementioned U-shaped structure along the C2 walls delineates B0 from the ambient gas. The eastern B0 wall shows higher abundances of SiO and HCN than the western wall \citep{GomezRuiz2013}, perhaps suggesting a pseudo-shock event in the region. 
There is also a ridge of emission connecting the east and west walls of B0, showing similar abundances to the western wall.

High spatial resolution, interferometric observations of the outflow reveal a clumpy structure within L1157 \citep{Benedettini2007}. Within all three regions, molecules are seen to peak in clumps $\sim$10-15\arcsec\, in diameter. In B1 and B2, these clumps lie within the shocked regions. In B0, these clumps are well correlated with the eastern wall, western wall, and the emission ridge.

\subsubsection{Structure Revealed by CARMA Maps}
\begin{figure*}
\includegraphics[width=\textwidth]{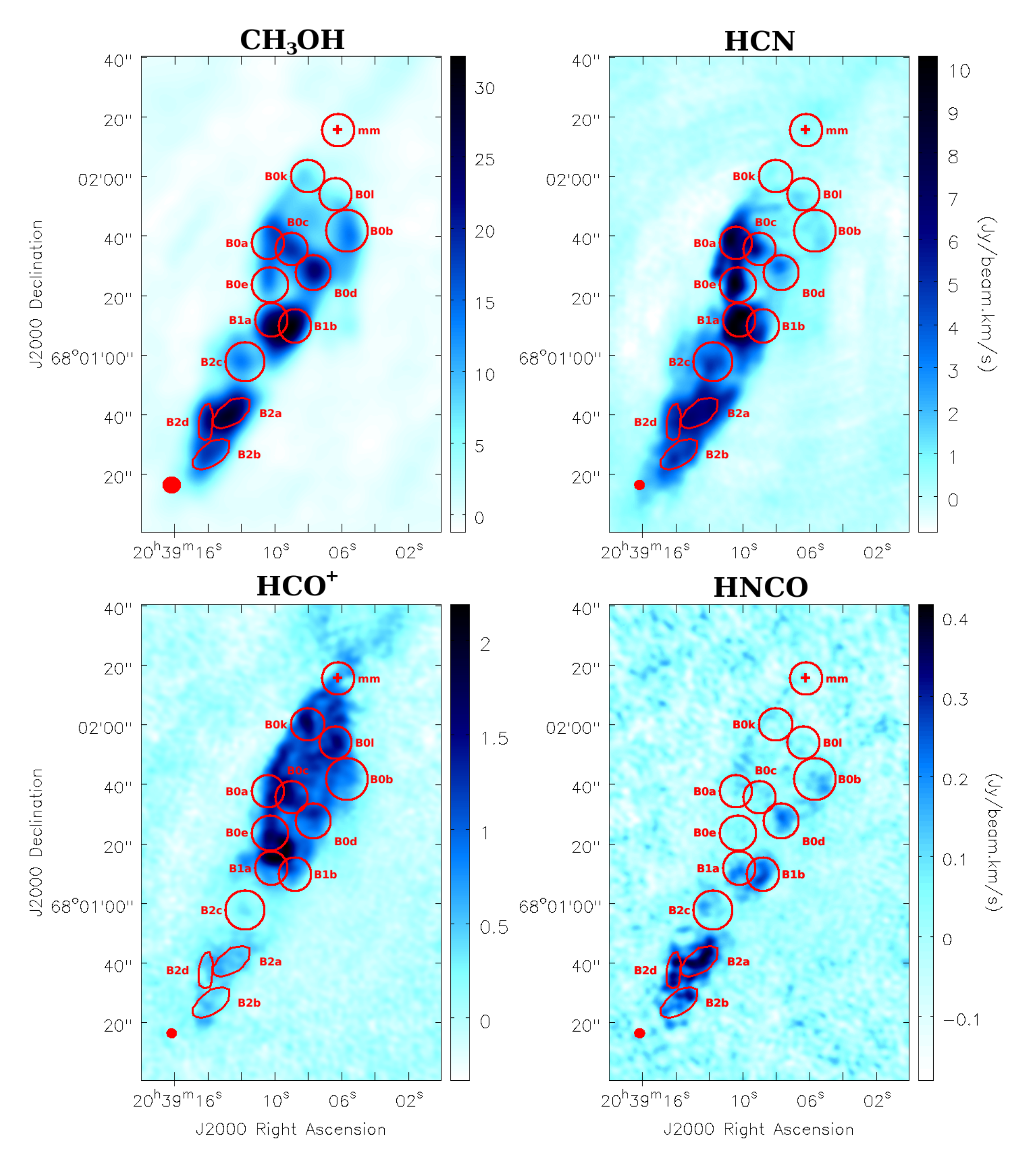} 
\caption{Integrated emission maps of CH$_3$OH ($2_k-1_k$), HCN (1-0), HCO$^+$ (1-0), and HNCO ($4_{0,4}-3_{0,3}$) with the clump regions and stellar source shown in red. The velocity ranges of integration were (-17.8 to $+$14.0 km s$^{-1}$), (-22.1 to $+$15.0 km s$^{-1}$), (-20.1 to $+$11.9 km s$^{-1}$), and (-6.45 to $+$6.9 km s$^{-1}$), respectively. The synthesized beam sizes are displayed as filled red in the lower left.}
\label{fig:all_molec_maps}
\end{figure*}

The HCN and CH$_3$OH transitions presented here were previously imaged by \citet{Benedettini2007} with similar resolutions. Indeed, the structure observed by CARMA agrees with the PdBI images previously published. While discussing the structure in this work, we therefore adopt the naming convention of \citet{Benedettini2007} and further added to by \citet{Codella2009} and \citet{GomezRuiz2013}. They defined 10 clumps in B0 (B0a-B0j), 8 clumps in B1 (B1a-B1h), and 3 clumps in B2 (B2a-B2c), based primarily on emission from CH$_3$OH, HCN, CH$_3$CN, SiO, and H$_2$CO. To maintain consistency with previous work, we specify elliptical regions corresponding to the clump positions reported in these papers. While the regions were described as ellipses $\sim$10-15\arcsec\, in size \citep{Benedettini2007}, we specify the dimensions of the ellipses with major and minor axis lengths determined based on the distributions observed in our images. We also define three new clumps: B0k and B0$\ell$ using the HCO$^+$ emission, and B2d using the HNCO emission. The coordinates and dimensions of the regions used are defined in Table \ref{tab:clumps_regions}.\added{ Also in Table \ref{tab:clumps_regions}, we list the region used to extract the spectra toward the gas surrounding L1157-mm, whose position was determined by \citet{Gueth1997}. While the source size is likely much smaller than the region used, the size and shape were chosen for comparison to clumps in outflow.} Figure \ref{fig:all_molec_maps} shows the integrated line emission of CH$_3$OH 2$_k$-1$_k$, HCN (1-0), HCO$^+$ (1-0), and HNCO ($4_{0,4}-3_{0,3}$), with the elliptical regions overlaid.

\begin{deluxetable}{l c c c c c c c}
\tablewidth{0pt}
\tablecaption{\added{Region} Positions and Sizes}
\tablecolumns{8}
\tablehead{
	\colhead{}& \multicolumn{3}{c}{Ellipse Center} & \colhead{} & \multicolumn{3}{c}{Ellipse Parameters} \\
	\cline{2-4} \cline{6-8} \\
	\colhead{\added{Region}} & \colhead{RA (J2000)} & \colhead{} & \colhead{Dec. (J2000)} & \colhead{} & \colhead{RA} & \colhead{Dec.} & PA \\
	& ($^{h\ \ \ }$ $^{m\ \ \ }$ $^{s}$) & \colhead{} & ($^{\circ\ \ \ }$ \arcmin$^{\ \ \ }$ \arcsec) & \colhead{} & (\arcsec)& (\arcsec) & ($^{\circ}$)
}
\startdata
B0a & 20 \ 39 \ 10.4 & & 68 \ 01 \ 38.0 & & 11 & 11 &-\\
B0b & 20 \ 39 \ 05.7 & & 68 \ 01 \ 42.0 & & 14 & 14 &-\\
B0c & 20 \ 39 \ 09.0 & & 68 \ 01 \ 36.0 & & 11 & 11 &-\\
B0d & 20 \ 39 \ 07.7 & & 68 \ 01 \ 28.0 & & 12 & 12 &-\\
B0e & 20 \ 39 \ 10.3 & & 68 \ 01 \ 24.0 & & 12 & 12 &-\\
B0k & 20 \ 39 \ 08.1 & & 68 \ 02 \ 00.4 & & 11 & 11 &-\\
B0$\ell$ & 20 \ 39 \ 06.4 & & 68 \ 01 \ 54.3 & & 11 & 11 &-\\
B1a & 20 \ 39 \ 10.2 & & 68 \ 01 \ 12.0 & & 11 & 11 &-\\
B1b & 20 \ 39 \ 08.8 & & 68 \ 01 \ 10.0 & & 11 & 11 &-\\
B2a & 20 \ 39 \ 12.6 & & 68 \ 00 \ 40.7 & & 7.5 & 15.5& 50\\
B2b & 20 \ 39 \ 13.8 & & 68 \ 00 \ 26.8 & & 7.5 & 15.5& 50\\
B2c & 20 \ 39 \ 11.8 & & 68 \ 00 \ 58.0 & & 13 & 13 &-\\
B2d & 20 \ 39 \ 14.1 & & 68 \ 00 \ 37.7 & & 5 & 12 & 0 \\
mm  & 20 \ 39 \ 06.2 & & 68 \ 02 \ 15.9 & & 11 & 11 & -
\enddata
\tablecomments{Regions based off of \citet{Benedettini2007} and \citet{Rodriguez2010}, with several new clumps assigned (B0k, B0$\ell$, and B2d). For non-circular regions, the semi-minor axis, semi-major axis, and position angle of the semi-major axis from straight north are given in that order. L1157-mm source size is likely much smaller, but\added{ the} region size\added{ and shape were} chosen for comparison to clumps in outflow.}
\label{tab:clumps_regions}
\end{deluxetable}

\subsubsection{CH$_3$OH}
CH$_3$OH 2$_k$-1$_k$ displays a clumpy structure that is consistent with the previous observations by \citet{Benedettini2007}. In our newly-assigned clumps of B0k, B0$\ell$, and B2a, we detect emission, although the clumps are not as prominent in CH$_3$OH emission as in HNCO and HCO$^+$. As has been found in previous observations, CH$_3$OH emission is generally observed to be brighter on the western side of B1, and has significant emission along the ridge in B0.

\subsubsection{HCN}
HCN (1-0) shows structure similar to CH$_3$OH, with significant clumping evident throughout L1157B.
However, whereas CH$_3$OH produces stronger emission on the western wall of the C2 cavity (i.e. in B1b, B0d, B0b), HCN emission is strongest on the eastern wall of C2 (i.e. in B1a, B0e, and B0a). This trend was also reported in \citet{Benedettini2007}. Significant structure is also seen outside of these clumps that is not seen in other molecules, especially toward the eastern wall of B0, with an arm protruding to the east of B0. This structure has been observed previously in SiO (5-4) as an ``arc-like'' structure to the east of the U-shape of C2 \citep{GomezRuiz2013}. HCN (1-0) has stronger emission toward the eastern clumps of B1, unlike CH$_3$OH, which is consistent with the chemical segregation previously observed \citep{Benedettini2007}. In B2, we observe significant emission toward B2a, B2b, and B2d.

\subsubsection{HCO$^+$}
HCO$^+$ (1-0) exhibits a more extended distribution, and is detected beyond the combined 10m-6m primary beam ($\sim$\,2\arcmin) of CARMA along the outflow axis. The HCO$^+$ emission is observed further into B0, tracing the C2 cavity, than the other molecules observed, and extending past the stellar source. We define two new clumps with HCO$^+$, B0k and B0$\ell$, located in the northern regions near the stellar source, L1157-mm. Strong emission is present in C2, particularly toward the eastern wall and the northern clumps, while much weaker emission is observed toward C1. The distribution is consistent with previous interferometric observations of HCO$^+$, where the majority of the emission was observed to trace the low-velocity ambient gas in B0 \citep{Bachiller2001,Choi1999}.

\subsubsection{HNCO}
HNCO exhibits a distribution distinct from the other observed species, with significantly-stronger line emission from B2.  
Weaker line emission is associated with B1 and the western wall of B0, similar to CH$_3$OH and unlike HCN. Previous single-dish observations have detected strong HNCO line emission toward B1 and B2, which are in agreement with our observations \citep{Rodriguez2010,Mendoza2014}. 

\subsection{Spectra and Column Density Determination}
\label{results:spectra&cd}
\begin{figure*}
\includegraphics[width=\textwidth]{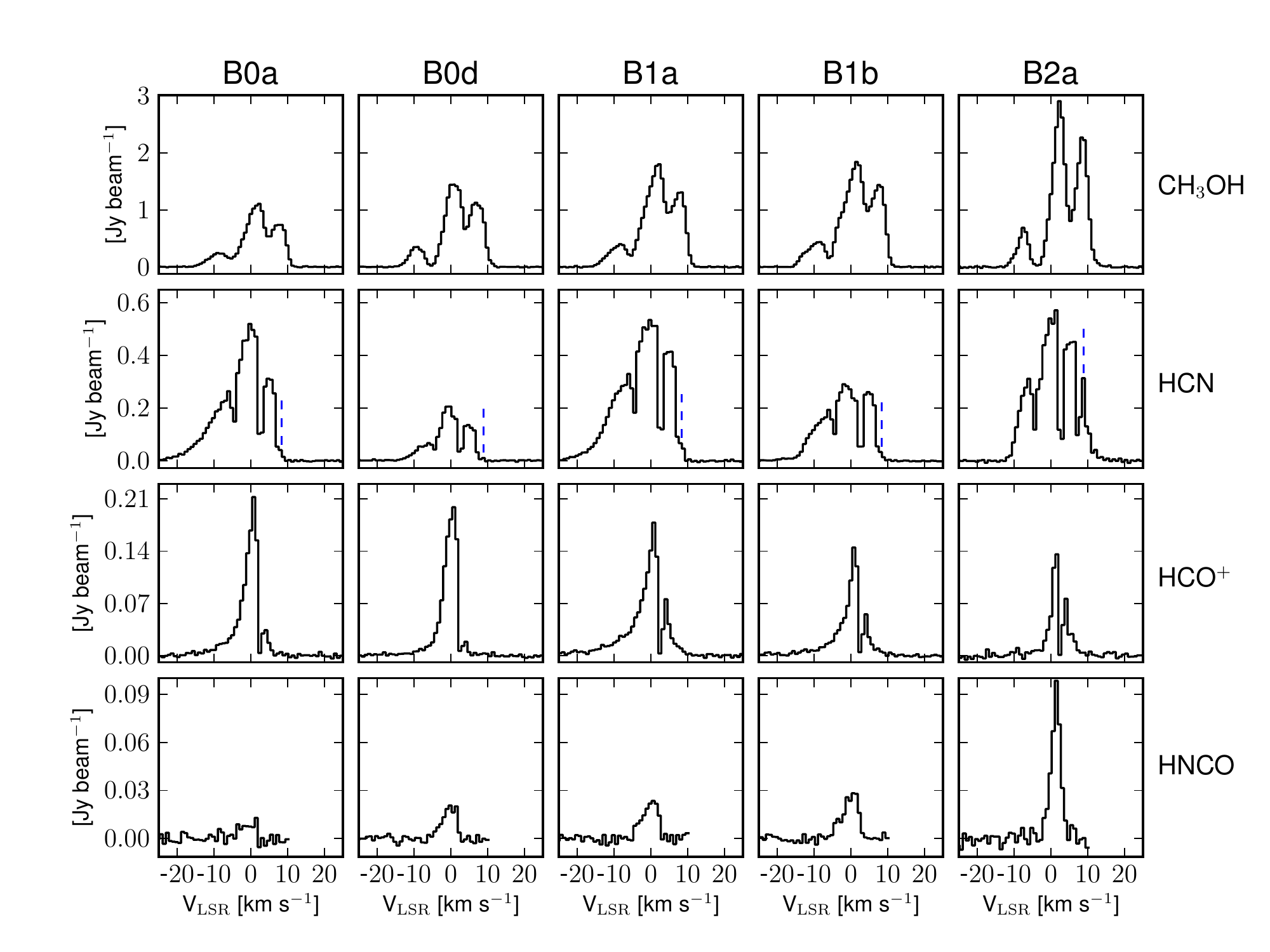} 
\caption{The extracted spectra from the regions B0a, B0d, B1a, B1b, and B2a are shown in black. For molecules with multiple transitions, CH$_3$OH and HCN, the velocities displayed are relative to the rest frequency of the central transition.\added{ For HCN, the unknown feature discussed in Section \ref{spectra:hcn} is marked with a blue dashed line. Not shown is the transition CH$_3$OH 2$_{1,1}$-1$_{1,0}$.}}
\label{fig:full_grid_spectra}
\end{figure*}

We extracted spectra from the regions defined in Table \ref{tab:clumps_regions}. Transitions of CH$_3$OH, HNCO, and HCN, including hyperfine components, were fit with Gaussian profiles, utilizing a modified version of the line fitting code described in \citet{Corby2015}. Figure \ref{fig:full_grid_spectra} displays the extracted spectra of the observed molecules at 5 representative regions which probe the emission at a variety of environments across L1157B: B0a, B0d, B1a, B1b, and B2a, with line fits overlaid. Spectra for all of the defined regions in Table \ref{tab:clumps_regions} are included in the Appendix \ref{appendix:spectra} and Gaussian fits parameters are in Appendix \ref{appendix:fits}.

An inspection of the line fit parameters reveal the kinematic structure of the outflow. For the purposes of studying the systematic velocities, we use the average velocities for the three strongest components of CH$_3$OH and the $F=1-1$ hyperfine transition of HCN, which will be further discussed in Sections \ref{spectra:ch3oh} and \ref{spectra:hcn}, respectively. \added{The gas surrounding the stellar source, L1157-mm, was found to have velocities between $v_{\text{LSR}}$\,$\sim$2.3-3.1 km s$^{-1}$}. For reference, the cloud velocity for this region has been found to be $v_{\text{LSR}}\sim$2.6 km s$^{-1}$ \citep{BachillerPerezGutierrez1997}.

In general, the clumps located in B0 south of the ridge, B1, and B2c are found to be kinematically related, with average systematic velocities around 0 to 1.5 km s$^{-1}$. Meanwhile, in the northern part of B0, the systematic velocities generally increase for each species: $\sim$1 km s$^{-1}$ for CH$_3$OH, $\sim$1.3 km s$^{-1}$ for HCN, and $\sim$1.6 km s$^{-1}$ for HNCO. Toward B2, the velocities are very consistent and larger than what was found in B1, southern B0, and B2c: $\sim$2 km s$^{-1}$ for CH$_3$OH, $\sim$0.9 km s$^{-1}$ for HCN, and $\sim$1.8 km s$^{-1}$ for HNCO. While the line widths of CH$_3$OH consistently remain around 3.4 km s$^{-1}$ throughout L1157B, HCN has smaller line widths in northern B0 ($\sim$2.1 km s$^{-1}$) than the rest of the L1157B, which displays similar line widths to CH$_3$OH of $\sim$3.5 km s$^{-1}$. Even more dramatically, the line widths of HNCO within these three kinematic regions are found to be distinct: $\sim$1 km s$^{-1}$ in northern B0, $\sim$5 km s$^{-1}$ in southern B0 and in B1, and $\sim$3 km s$^{-1}$ in B2. Curiously, while the HNCO (4-3) line velocity shift in B2c is more related to the apex of C2, the line width is closer to the rest of B2. It is important to note that the current naming scheme for clumps in L1157 does not appear to accurately reflect the regions that are kinematically associated. Based on these findings, HNCO may prove to be a useful probe for disentangling kinematic structure in these sorts of regions, perhaps due to its unique formation pathway, as discussed in Section \ref{discussion:hnco}.

To compute the column densities of CH$_3$OH, HCN, and HNCO, we utilize the non-LTE radiative transfer code, {\sc radex} \citep{VanderTak2007}. As inputs of the gas temperatures, we adopt the physical conditions determined in Paper I, using a \textsc{radex} analysis of $\sim$30 transitions of CH$_3$OH observed with the CSO. As the CSO observations only included single pointings toward B1, encompassing B1a, B1b, and B0d, and toward B2, containing B2a, B2b, and B2d, we only computed the column densities in these six regions. While a similar analysis could be performed for the remainder of B0, previous estimates for the kinetic temperatures and gas densities throughout B0 have not been consistent in the literature \citep{Bachiller2001,Choi1999}. It should be noted that further observations could help constrain the physical environment in B0, which would provide contrast to B1 and B2 shocks.

We apply derived parameters of $T_k=60$ K, $n_{\text{H}_2}=3\times10^5$ cm$^{-3}$ toward B1 and $T_k=50$ K, $n_{\text{H}_2}=6\times10^5$ cm$^{-3}$ toward B2. While Paper I used a 2-component fit with an extended, cool component and a compact, warm component, we only adopt the warm component parameters, as the warm component should \added{dominate} the emission originating from these small clumps. With these input conditions and line parameters from the Gaussian fits to the spectra, we obtain column densities reported in Table \ref{tab:molecules_abundances}. 

For the purposes of this analysis, the \added{spectral lines} without anomalous features, such as non-symmetric broadening, significant self-absorption, and line-blending, were used to constrain the abundances. The individual cases are discussed below. The derived column densities of each species are given in Table \ref{tab:molecules_abundances}. Liberal uncertainties are assumed due to the small number of transitions observed, asymmetric line profiles, and the inherit uncertainties discussed in Paper I to be $\sim$32\%.

\begin{deluxetable}{l c c| c c c| c c c}
\tabletypesize{\footnotesize}
\tablewidth{0pt}
\tablecaption{Molecular Abundances and Enhancements}
\tablehead{
	\colhead{} &  \multicolumn{2}{c}{CH$_3$OH} & \multicolumn{3}{c}{HCN} &  \multicolumn{3}{c}{HNCO} \\ 
	\colhead{ } & \colhead{$N$} & \colhead{$\frac{N}{N_{\text{H}_2}}$} &\colhead{$N$} & \colhead{$\frac{N}{N_{\text{CH}_3\text{OH}}}$} & \colhead{$\frac{N}{N_{\text{H}_2}}$} &\colhead{$N$} & \colhead{$\frac{N}{N_{\text{CH}_3\text{OH}}}$} & \colhead{$\frac{N}{N_{\text{H}_2}}$} \vspace{0.1cm} \\ 
	& ($10^{15}$\,cm$^{-2}$) & ($10^{-6}$) & ($10^{13}$\,cm$^{-2}$) & ($10^{-2}$) & $(10^{-8})$ & ($10^{13}$\,cm$^{-2}$) & ($10^{-2}$) & $(10^{-8})$
}
\startdata 
B0d & 1.5 &1.5 & 2.6 & 1.7 &2.6 & 2.5 & 1.6 &2.5 \\
B1a & 1.9 &1.9 & 10. & 5.3 &10. & 3.2 & 1.7 &3.2 \\
B1b & 2.1 &2.1 & 5.3 & 2.5 &5.3 & 3.8 & 1.8 &3.8 \\
B2a & 2.7 &2.7 & 9.9 & 3.7 &9.9 & 8.0 & 3.0 &8.0 \\
B2b & 1.6 &1.6 & 6.6 & 4.2 &6.6 & 5.5 & 3.5 &5.5 \\
B2d & 2.1 &2.0 & 8.0 & 3.9 &7.9 & 7.5 & 3.6 &7.5 
\enddata
\tablecomments{For the three molecules studied in depth, the column densities calculated through \textsc{radex} are given at each region that coincides with Paper I CSO pointings, along with the enhancement relative to CH$_3$OH and overall abundance relative to H$_2$. Errors, as described in Paper I, are calculated to be $\sim$32\%}
\label{tab:molecules_abundances}
\end{deluxetable}

We further convert from column densities to abundances with respect to hydrogen. \citet{BachillerPerezGutierrez1997} determined that $N_{\text{H}_2}\sim10^{21}$ cm$^{-2}$ in B1 and $N_{\text{H}_2}\sim5\times10^{20}$ cm$^{-2}$ in B2 using the optically thick $^{12}$CO line emission and conversion by $N_{\text{H}_2} \simeq 10^4 N_{\text{CO}}$. However, using measurements on the optically thin isotopologues of CO, such as $^{13}$CO, higher column densities are derived. Additionally, multiple authors have suggested that a higher conversion factor of $N_{\text{H}_2} \simeq 3\times 10^4 N_{\text{CO}}$ may be more appropriate \citep{Bolatto2013, Lefloch2012}. Given these ambiguities, we adopt $N_{\text{H}_2}\sim2\times10^{21}$ cm$^{-2}$ toward B1 and $N_{\text{H}_2}\sim10^{21}$ cm$^{-2}$ toward B2 based on $N_{\text{CO}}\sim 9\times10^{16}$ cm$^{-2}$ in B1 \citep{Lefloch2012} and $N_{\text{CO}}\sim 5\times10^{16}$ cm$^{-2}$ in B2 \citep{BachillerPerezGutierrez1997}..

Furthermore, roughly half of the CO line flux towards B1 arises within a blueshifted wing component in the velocity range of -15 to -4 km s$^{-1}$, previously designated the \textit{g2} component \citep{Lefloch2012}. This wing component is evident in HCN and HCO$^+$, but is absent from the profiles of CH$_3$OH and HNCO, as seen in Figures \ref{fig:full_grid_spectra} and \ref{fig:blue_wing}. Similar effects have been described by \citet{Tafalla2010}, but with their designated f-wing to range over -7.5 to -3.5 km s$^{-1}$, as determined by the transition from optically thin to thick through the $^{13}$CO(2-1) to CO(2-1) ratio at each given velocity. Because CH$_3$OH and HNCO do not exhibit this wing component, we focus on the primary velocity peak, which is more Gaussian shaped. Taking this feature into account, we estimate that $N_{\text{H}_2}\sim10^{21}$ cm$^{-2}$ in the primary component toward B1, in the \textit{g2}-wing component of B1, and toward B2, in agreement with the derived hydrogen column within the slow wing of \citet{Tafalla2010}. We assume that this hydrogen column density is the same towards all six clumps. As is typical of abundance measurements, the reported abundances are therefore approximate, and relative abundances determined from lines within this dataset are more precise. Below, we briefly discuss the spectrum of each molecule and observed patterns in column density and abundance.

\subsubsection{CH$_3$OH}
\label{spectra:ch3oh}
We detect the four $2_k-1_k$ transitions of CH$_3$OH in all of the clumps used in this study, and we detect weak CH$_3$OH emission towards the stellar source, consistent with previous observations \citep{GomezRuiz2013}. The relative intensities of the three stronger lines ($2_{-12}-1_{-11}$, $2_{02}-1_{01}$++, $2_{02}-1_{01}$) remained a very consistent ratio of 4:3:1, respectively, throughout the regions.
We note the presence of a small wing on the blue shifted side of the CH$_3$OH line profiles towards B1a and b and towards B0a, c, d, and e. This wing component spans a velocity range of -4 to -2 km s$^{-1}$, which is distinct from the $g2$ wing component which extends to -15 km s$^{-1}$. This has been observed previously by \citet{GomezRuiz2013}. 
As was \added{also} reported by \citet{GomezRuiz2013}, CH$_3$OH emission in L1157B is detected at velocities only slightly blueshifted \added{($\left|\Delta v_{\text{LSR}}\right|<3.5$ km s$^{-1}$)} to the \added{cloud} velocity, with generally larger offsets observed at distances further from L1157-mm (Table \ref{tab:spectral_fit_ch3oh}). There is also a small, blue wing with a width of $\sim$4 km s$^{-1}$ observed towards B1 and B0a, c, d, and e. Since B1 is located at the apex of the bow shock, this wing may originate from the terminal velocity of the shock as it propagates into cold, dense, unshocked material.

The reported column density for CH$_3$OH is the SNR-weighted average of the individual column densities derived per transition using \textsc{radex}. The column densities remained fairly consistent for the various regions, being on the order of $\sim$1.5-2.7$\times10^{15}$ cm$^{-2}$. Similarly, the abundances relative to hydrogen are estimated to be $\sim$1.5-2.7$\times10^{-6}$, with the highest abundances toward B2a. These are displayed in Table \ref{tab:molecules_abundances}. These values are consistent with the results of Paper I, which obtained column densities of $\sim$3$\times10^{15}$ cm$^{-2}$ from CSO observations towards B1 and B2. The small differences in column densities may be attributed to Paper I's use of a two-component fit in the \textsc{radex} analysis, as well as some of the extended emission from CH$_3$OH in the cooler component being resolved out by CARMA's beam. Given the general agreement with previous observations, we will study the abundance enhancement of other observed molecules relative to methanol.

\subsubsection{HCN}
\label{spectra:hcn}
HCN has been detected in all clumps studied in L1157B, as well as the stellar source. Velocity offsets of HCN (1-0) were observed to be slightly blueshifted to the systemic velocity, but with no significant gradient over distance from L1157-mm observed. As seen in Figure \ref{fig:full_grid_spectra}, the three hyperfine components are clearly resolved, and we report two anomalous features. 

\begin{figure*}
\includegraphics[width=\textwidth]{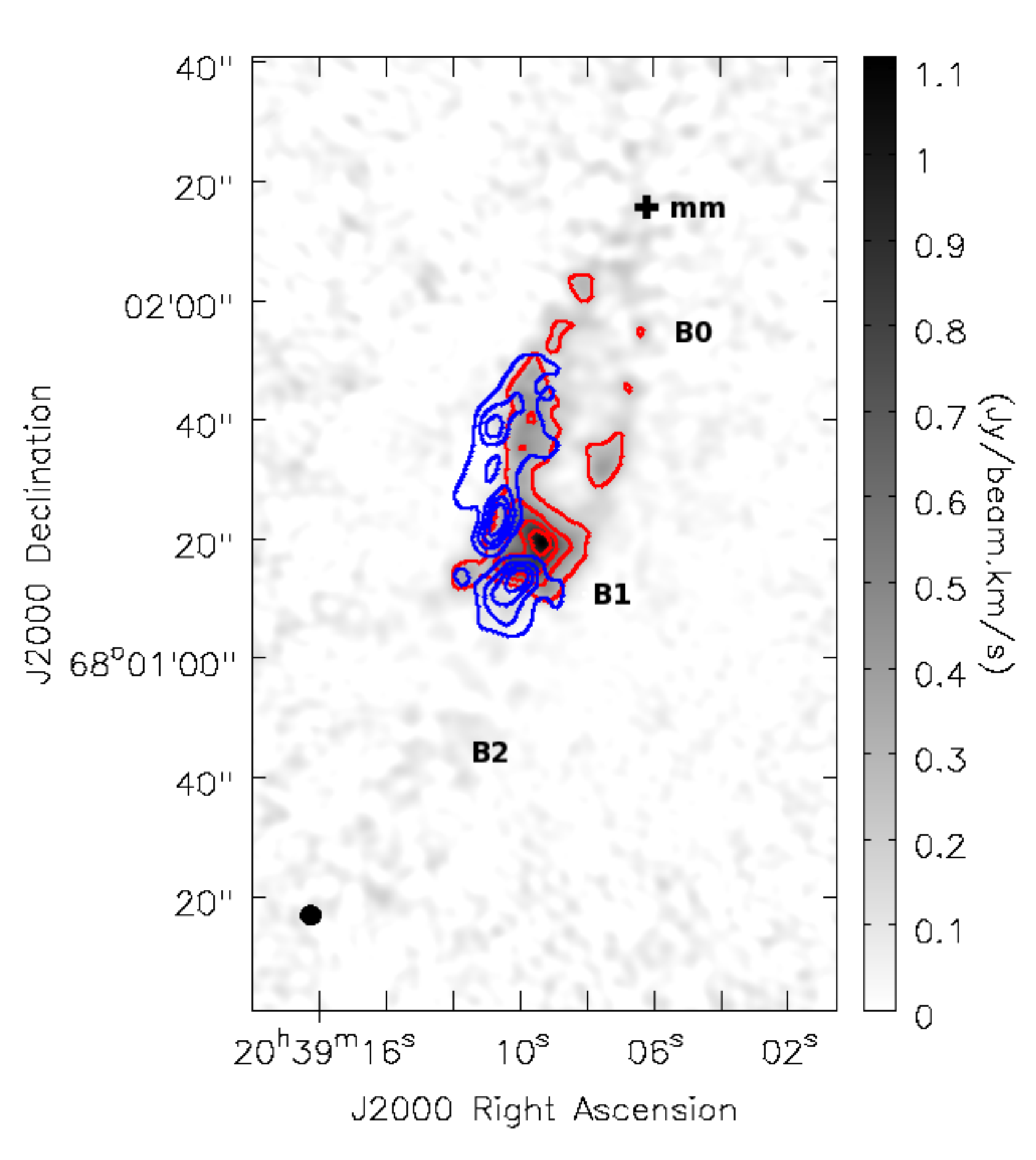} 
\caption{Integrated emission maps of the blue wing component of HCN (1-0) (blue contours at levels of 0.5, 1.0, 1.5, \& 2.0 Jy beam$^{-1}$ km s$^{-1}$) and HCO$^+$ (1-0) (raster, red contours at levels of 0.22, 0.45, 0.67, \& 0.89 Jy beam$^{-1}$). The velocity ranges of integration were both -15 to $+$4 km s$^{-1}$. The synthesized beam size is displayed as filled black in the lower left.}
\label{fig:blue_wing}
\end{figure*}

First, the $F=0-1$ hyperfine component is seen to be broadened and contains a prominent blueshifted wing toward most clumps in B0 and B1, as seen in previous observations of this line \citep{Benedettini2007}. 
This component is associated with the $g2$ component observed in CO \citep{Lefloch2012} as well as in HCO$^+$ in our observations. Here, the lower-frequency edge of the blue wing was set by where the morphology of the line emission significantly changed before the peak of the $F=0-1$ transition. The integrated HCN (1-0) emission in the velocity range of -15 to -4 km s$^{-1}$, with respect to the $F=0-1$ hyperfine component, is shown in Figure \ref{fig:blue_wing} overlaid on HCO$^+$ emission within this velocity range. 
This velocity component clearly maps to the apex and eastern wall of C2 and is notably absent toward B2.
This may imply that the wing emission originates from the shock as it propagates through the semi-processed material between C1 and C2, and thus has not been significantly slowed down. This is further evidenced by SiO emission, which displays broader velocity wings ($\sim$20 km s$^{-1}$) \citep{GomezRuiz2013}. This will be further discussed later. 

\begin{figure*}
\includegraphics[width=\textwidth]{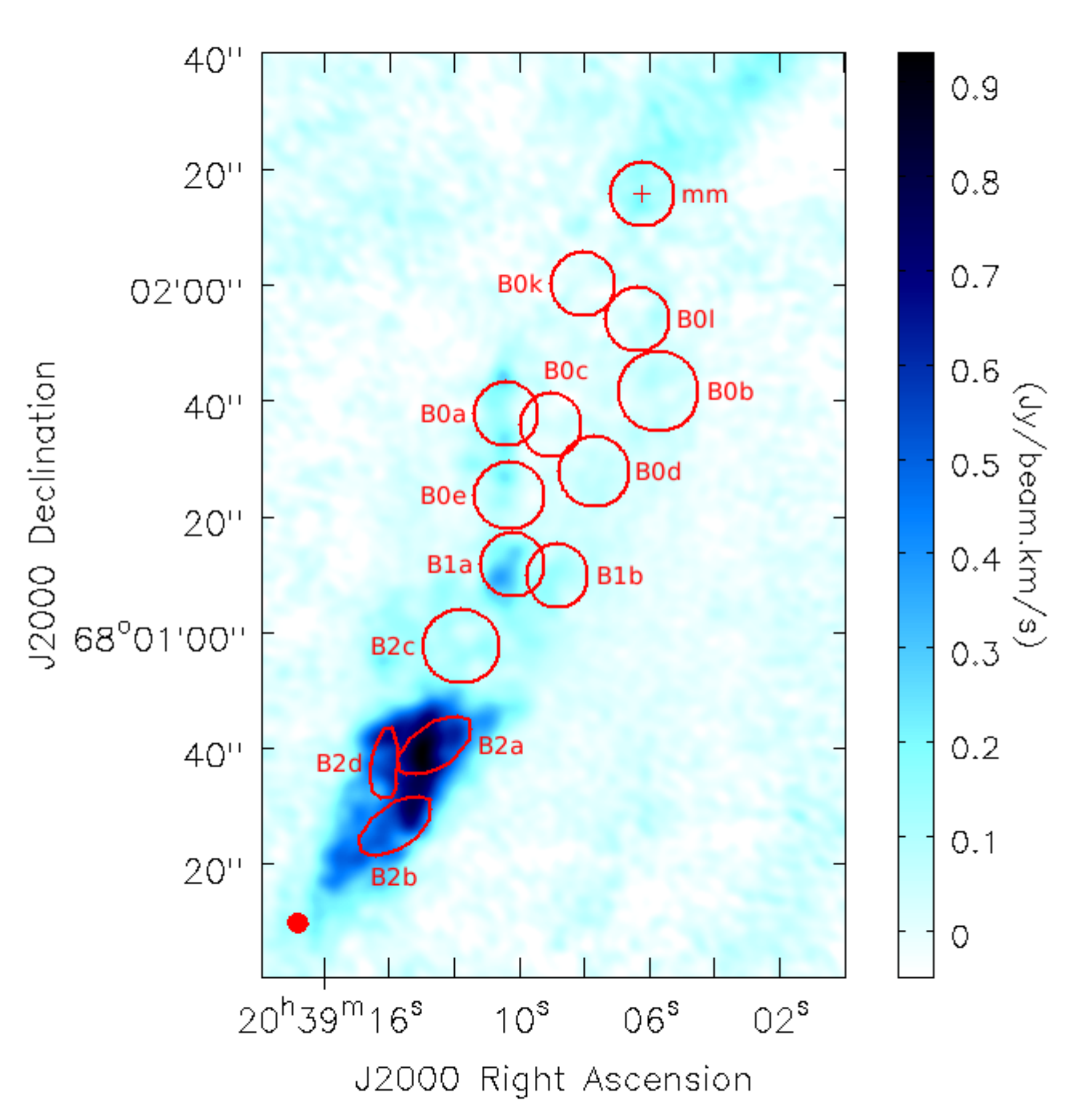} 
\caption{Integrated emission maps of the 4$^{\text{th}}$ unknown component near HCN (1-0), with the regions defined in Table \ref{tab:clumps_regions} displayed in red, along with the stellar source. The synthesized beam size is displayed as filled red in the lower left.}
\label{fig:hcn_4th}
\end{figure*}

The second anomalous feature in the spectrum is a strong line at $\sim$4 km s$^{-1}$, with respect to the $F=1-1$ transition, toward B2a, B2b, and B2d. This can be seen in B2a of HCN in Figure \ref{fig:full_grid_spectra}. The integrated emission of this fourth component can be seen in Figure \ref{fig:hcn_4th} to be nearly exclusively in B2. The flux observed in B1a and the eastern wall of B0 is blended emission from the red tail of the $F=1-1$ line and not a separate component. This supposed fourth line could be explained by a second, prominent velocity component redshifted by approximately the hyperfine spacing. This would result in the central two observed features being a blend of two components, $F=2-1$/$F=0-1$ and $F=1-1$/$F=2-1$ for the second and third most blueshifted features, respectively. Meanwhile, the outer two features are unblended, single-transition features $F=0-1$ and $F=1-1$ for the first and fourth most blueshifted features, respectively. However, the relative emission strengths of the hyperfine components do not clearly indicate this is the case. Further, it is curious that a red-shifted component is not apparent in the spectral lines of other species observed in B2, and no similar feature has been reported in the literature for another molecule. But if the feature is a second velocity component, then this would provide further evidence that B2c is kinematically related to the apex of C2, and not to the rest of B2. Another possible explanation could be that this is an unidentified line of a separate species, which may explain why it is primarily in B2, similar to HNCO. However, it seems unlikely that we would observe an unknown transition of similar strength to HCN (1-0), with no reasonable candidate transitions within the NRAO Splatalogue database\footnote{Available at \url{www.splatalogue.net} \citep{Remijan2007}}. Observation of the HCN (2-1) has the potential to clarify the scenario. 

Finally, the observed line intensity ratios of HCN were distinct from the optically thin, LTE hyperfine line ratios of $F = (2-1):(1-1):(0-1) = 5:3:1$. The observed ratios are about ${\sim1.76:1.35:1}$, respectively. While we do not expect LTE conditions, these line ratios cannot be accounted for by \textsc{radex} under physical conditions consistent with previous observations of L1157. This may indicate a pumping process producing weak masing, which should be accounted for by \textsc{radex}. It is also possible that, in part, the potential two blended velocity components discussed previously may cause some of the deviation from the thermal line ratios. Recent work by \citet{Mullins2016} on anomalous hyperfine line strengths also point to a strong dependency on optical depth in combination with gas infall or dynamics as a potential explanation of these observed ratios. 

Because the blue wing component appears to be associated with different gas than the clumps, we report HCN column densities from \textsc{radex} calculations solely on the $F=1-1$ transition. This hyperfine line should not include substantial flux contributed by the blue shifted wing. It has been previously shown that accurate modeling of HCN hyperfine transitions requires that radiative transfer calculations be explicitly applied to each $F$-level independently.  In particular, the $J=1-0$, $F= 0-1$ transition is prone to producing spurious results, especially when the line ratios are not well-described by a single excitation temperature \citep{Mullins2016}. The derived column densities are of order $10^{13}-10^{14}$ cm$^{-2}$, as shown in Table \ref{tab:molecules_abundances}, with higher column densities toward B1a and B2. The HCN to CH$_3$OH column density ratio varied between $1.7-5.3\times10^{-2}$. \citet{Benedettini2007} reported larger column densities by up to an order of magnitude in all regions except B0d. However, they derived their column densities assuming LTE, whereas we utilized the non-LTE code \textsc{radex} \citep{VanderTak2007}. The non-LTE approach is likely more accuate, especially given the non-thermal hyperfine ratios.

\citet{Benedettini2007} report that HCN (1-0) undergoes strong self-absorption in observations toward L1157B, which would result in an underestimation of their reported column density. It is interesting to note that toward all regions in L1157B the troughs between the hyperfine features lie very close to what would result from HCN residing in ambient gas (i.e. shifted by the systemic velocity of $\sim$2.7 km s$^{-1}$). The red side of each of these features also appears to fall off with slopes that are not indicative of a typical Gaussian profile, suggesting the line may be significantly self-absorbed. A possible analogous absorption feature is also observed in the HCO$^+$ (1-0) emission. It does not appear that CH$_3$OH or HNCO have similar self-absorption features.

{If the observed profiles for HCN are indeed due to self-absorption, the abundances derived here would be underestimated. As a result, the molecular enhancements would be even larger than those determined here. We do not have sufficient data to make a claim as to the likelihood that this is self-absorption, but we suggest this may be an intriguing avenue for follow-up study.


\subsubsection{HCO+}
\label{spectra:hco+}
HCO$^+$ has been detected toward each region in L1157B and in the stellar source. As seen in Figure \ref{fig:full_grid_spectra}, the emission displays significant self-absorption toward all positions in the outflow at a velocity of 3$-$4 km s$^{-1}$. However, this self-absorption is not detected towards the stellar source. Due to the non-Gaussian line shapes and the self-absorption, we do not fit the spectra or determine HCO$^+$ column densities here.

The blueshifted $g2$ wing of HCO$^{+}$ (1-0) is observed, spanning a similar velocity range of HCN (1-0). As shown in Figure \ref{fig:blue_wing}, both of these blueshifted components map to the apex and eastern wall of C2, implying that the source of these emission wings may originate from the same physical structure. Figure \ref{fig:hco_2wing} shows blue wing ($v_{\text{LSR}}=$-15 to -3.86 km s$^{-1}$) and red wing ($v_{\text{LSR}}=$4.3 to 10 km s$^{-1}$) of the HCO$^+$ emission along with the central peak ($v_{\text{LSR}}=$-3 to 3.5 km s$^{-1}$). The transition from the blue wing to the central peak was determined by where the morphology significantly changed, as discussed above, while the red wing/central peak boundary was the velocity where the self-absorption was strong enough to make the observed line strength effectively zero. Here, the central peak is observed to trace the ambient gas throughout the C1 cavity, with a slight peak along the ridge. The redshifted emission continues far past the stellar source, as would be expected. Interestingly, there are also redshifted components also observed at concentrated peaks along the apex of B0/1, B0$\ell$, and B2. While these are not necessarily separate components, it is possible that the emission originates from gas from the far side of the C1 and C2 cavity.

\begin{figure*}
\includegraphics[width=\textwidth]{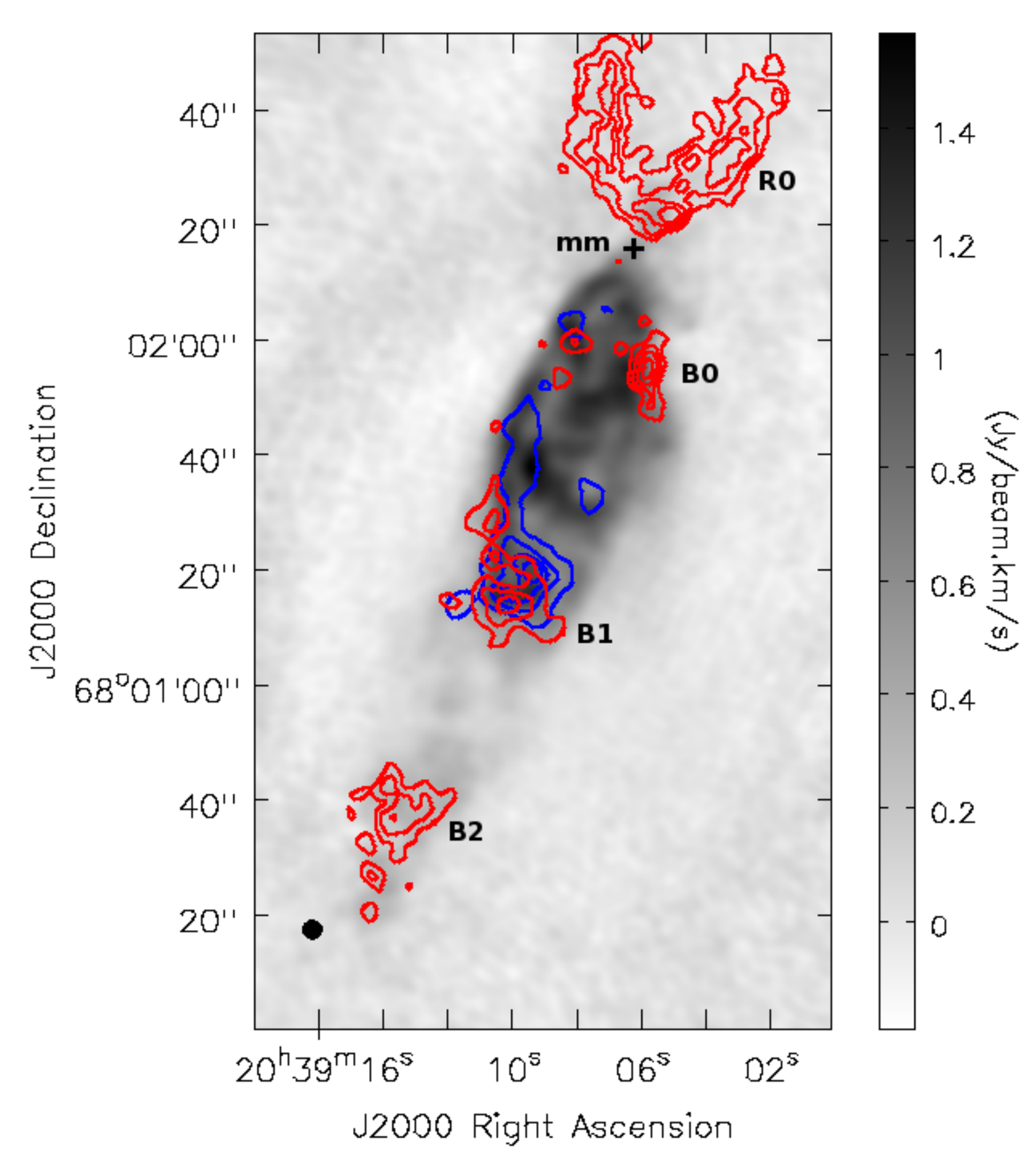} 
\caption{Integrated emission map of HCO$^+$ (1-0) with contours of its blue wing (-15 to -3.86 km s$^{-1}$ at levels of 0.18, 0.37, 0.55, \& 0.74 Jy beam$^{-1}$) and red wings ($+$4.3 to $+$10 km s$^{-1}$ at levels of 0.08, 0.16, 0.24, \& 0.33 Jy beam$^{-1}$) in their respective colors over the central peak raster (-3 to $+$3.5 km s$^{-1}$). The synthesized beam size is displayed as filled black in the lower left.}
\label{fig:hco_2wing}
\end{figure*}

\subsubsection{HNCO}
\label{spectra:hnco}
HNCO was detected toward all of the regions in L1157B, but not toward the stellar source.
Although weak, a wing component from -3 to -1 km s$^{-1}$ is apparent in the spectra towards B1 and the ridge of B0. This occupies the same velocity range and general spatial distribution as the weak wing component in CH$_3$OH, but is distinct from the $g2$ component evident in the spectra of HCN and HCO$^+$.
Our \textsc{radex} calculations based on our Gaussian fits determine the column densities of HNCO to vary between $\sim$2.5-8.0$\times10^{13}$ cm$^{-2}$ across the entire line, comparable to the measured column densities of HCN. A factor of 2 increase in column density between B2 and B1 is immediately apparent. This corresponds to the $N($HNCO$)/N($CH$_3$OH) increasing by a factor of 2 ($\sim$1.7$\times10^{-2}$ to $\sim$3.5$\times19^{-2}$) and similar for abundance enhancement relative to H$_2$, which is consistent with previous observations \citep{Mendoza2014}.

\subsubsection{CH$_3$CN}
\label{spectra:ch3cn}
A significant feature appears in three edge channels of one of our windows. Given a similar velocity profile as our other observed molecules, the transition appears to coincide with the $5_k - 4_k$ transition of CH$_3$CN.
The integrated emission of these three channels \added{display a distribution similar to CH$_3$OH, with significant emission observed throughout the major shocked regions of L1157B.}
Very little emission appears to come from any of the less prominent shocks and ambient gas. It has been detected previously in B1 in $14_k - 13_k$ \citep{Arce2008} and mapped in $8_k - 7_k$ \citep{Codella2009}, with a distribution consistent with our observations. These both report column densities between 10$^{12}$-10$^{13}$ cm$^{-2}$.

\section{Discussion}
\label{discussion}

A few trends emerge from the results. First, it is apparent that the northernmost clumps in B0 are kinematically distinct from the material in the south of the ridge. Further, the clumps in B1 and south of the ridge in B0 are more kinematically and structurally similar compared to northern B0. Second, a broad ($\sim$15 km s$^{-1}$) component, referred to as $g2$, has been observed in HCN and HCO$^+$, as well as CO and SiO in previous work \citep{Lefloch2012}. It is predominately located along the eastern wall of the C2 cavity and near the apex of C2, most significantly towards B0e. This component contributes a significant fraction of the molecular line emission by these species in the eastern portion of C2. Furthermore, we notice possible east-west differentiation within C2 and also between B1 and B2.

\subsection{East/West Chemical Contrast in C2} 
Previous literature has discussed the existence of a chemical differentiation within C2 \citep{Gueth1998,Tafalla1995,Bachiller2001,Benedettini2007,Benedettini2013}. While species including HCN and SiO produce strong line emission along the eastern wall of C2, other molecules such as CH$_3$OH and OCS are seen stronger along the western wall, especially towards the apex. We observe this trend in the species studied here as well.  In our CARMA images, CH$_3$OH is seen to be strongest at the western wall of B1 and along the ridge in B0, although this relative abundance enhancement is less significant compared to the other studied species. With the first interferrometric maps of HNCO presented here, it can be seen that this species is even more concentrated to the eastern wall of C2 than CH$_3$OH. Meanwhile, HCN is substantially stronger on the eastern wall compared to the western. As the $g2$ component preferentially contributes to the emission seen in the eastern wall, it is plausible that the apparent differentiation results from the contribution of this wing. However, considering only the emission by the HCN (1-0) $F=1-1$ transition, for which the wing component should not contribute substantially, it appears that the east-west differentiation still persists. As reported previously, the column density of CH$_3$OH is relatively consistent near the apex of C2, while a significant enhancement is observed for HCN in B1a over B1b \citep{Benedettini2013}. In our analysis, we found this to be an increase of column density of about a factor of two, which is a similar enhancement found along the eastern wall up to the ridge.  

\citet{Bachiller2001} assert such chemical segregation is due to chemical stratification within the cloud. \citet{Gueth1996} proposed that as the precessing jet sweeps from east to west, the western wall of C2 would be interacting with new, unshocked material in the molecular cloud. Meanwhile, the eastern wall is expanding into material already processed during the previous shock that produced C1.

In the western wall, especially towards the apex in B1, abundances would be indicative of protypical ice mantle desorption as the shock interacts with grains that previously resided in cool, quiescent gas. This results in an enhancement of grain-species, which we observe with CH$_3$OH and HNCO. On the other hand, the eastern wall is interacting with dust grains that have already been recently shocked by the B2 event. As a result, the ice mantles would not have had time to redeposit onto the surface of the grains. The shock would instead destroy the complex species that were brought into the gas phase during the first shock, and further erode the bare dust grain. The resulting products of these complex molecules would then be enhanced in these regions, as seen in HCN and HCO$^+$. This is supported by the reported anti-correlation of CN and HNCO in this region \citep{Rodriguez2010}, as CN is likely to be the product of this destruction. In addition, this also explains the distribution of SiO toward the eastern wall of C2, as it would be enhanced in the shocking of bare-grains compared to grains with significant ice-mantles.  In \citet{Benedettini2007}, emission from HC$_3$N, HCN, CS, NH$_3$, and SiO were 
found to be brighter in eastern clumps in C2, while CH$_3$OH, OCS, and $^{34}$SO emissions were more prominent in western clumps. The segregation of these specific species aligns with our proposed scenario.

This scenario is further supported by the velocity wings observed in these species. While the chemical differentiation within C2 is seen both in the wing and the primary velocity components, the wing components probe the properties of the shock that would affect the chemistry in the entire region. CH$_3$OH and HNCO are observed to have a small, $\sim$4 km s$^{-1}$ blue wing primarily towards B1b, as was previously discussed. This wing could be tracing the shock as it hits, and is thus slowed down, by the cold, dense, and unshocked gas. Meanwhile, HCO$^+$ and HCN are observed to have a large, blue, $g2$ wing primarily towards B1a and the eastern wall of C2. Here, the shock would not be slowed down because it is interacting with gas that has already been processed and recently shocked, and thus is less dense than towards B1b. This higher velocity scenario is also supported by the higher abundance of SiO towards the eastern wall, as it is produced more efficiently in high-velocity shocks, and where the dust grains are already bare due to the previous shock.

It should be noted that observations by \citet{GomezRuiz2013} of a pair of H$_2$CO lines with differences of excitation energies of $\Delta T\sim47$ K show significant\added{ly different} emission on opposite edges of B1, suggesting these asymmetries could instead be due to excitation effects. However, the majority of the lines studied here have relatively low excitations and still display chemical differentiation. A follow up study of higher energy transitions of these species would help determine the significance of excitation on the chemical differentiation in C2. Recent observations of isotopologues of CO by \citet{Kwon2015} may suggest that there are in fact two bipolar jets, each of which could contribute to a different wall of C2. While this may explain the asymmetric brightness of certain lines, this does not necessarily explain the distinct chemical segregation within the region.   

\subsection{Enhancements in B2}
Among the species studied, abundances toward the shocked regions in B2 were found to be either equivalent or higher compared to the B1 shocks. The abundance relative to H$_2$ of HCN toward B2a and B1a are nearly equivalent. For CH$_3$OH, B2 appears to be enhanced, although it is only just outside of our reported uncertainties. And in HNCO, we see a factor of two increase in abundance relative to CH$_3$OH. Meanwhile, HCO$^+$ emission is significantly weaker in B2 relative to B0 and B1. To determine what may cause this \added{difference}, it is necessary to determine how B1 and B2 vary.

Since the shock in B2 is older than in B1, the observed abundances could be a result of viewing changes on a chemically-relevant timescale. If this proves significant, then it would imply that the initial liberation of molecules from the grain by the shock may not be the only prominent process that occurs in these environments. Instead, post-shock, gas-phase reactions would become significant, as these regions are at higher densities and temperatures than the pre-shock environments. \added{E}xamples of these will be discussed briefly in Section \ref{sec:formationchem}.

\citet{GomezRuiz2013} have also proposed that the shock in B2 is $\sim$10 km s$^{-1}$ faster than B1 from observations of SiO and its line profile. Higher shock velocities can result in more efficient erosion of ices on grain surfaces  and the mantles themselves, as \added{has} been shown in models of sputtering on grains. \citep{JimenezSerra2008,VanLoo2013}. If this difference in shock velocities is significant to the proceeding chemistry, then the higher abundances of CH$_3$OH and HNCO could be explained by a higher liberation rate from the grain. 

\subsection{Formation Chemistry Implications}
\label{sec:formationchem}

\subsubsection{CH$_3$OH}
\label{discussion:ch3oh}
Overall, the consistent abundance of CH$_3$OH indicates that its primary source is from the liberation of grains during the shock. The slightly higher abundances toward the relatively cooler, older shock, B2, may imply that the CH$_3$OH may have a production pathway as the young shock-tracing molecules react following their initial release into the gas phase. Chemical models of outflows have shown that methanol abundance is elevated in higher gas density regions, from $10^5$ to $10^6$ cm$^{-3}$ \citep{Viti2004}.  This difference in density was observed when comparing the gas densities of B1 and B2 from Paper I, supporting this interpretation. \citet{Rawlings2013} has discussed that recently shocked material surrounding the grains is at high enough densities that three-body reactions may occur efficiently, such as with CH$_3$, OH, and H$_2$O (as a passive third body) to produce CH$_3$OH. It is also possible that the larger shock velocity in B2 relative to B1, as proposed by \citet{GomezRuiz2013}, could result in a more efficient liberation of grain species.

\subsubsection{HCN}
\label{discussion:hcn}
If the scenario proposed for the east/west chemical differentiation is valid, then the formation of HCN would be significantly impacted by the formation of CN from the destruction of more complex species in the gas phase. Following the enhancement of CN in the region, HCN could be readily produced as well. Similar iterations of this interpretation have been proposed by \citet{Benedettini2007}.

\subsubsection{HCO$^+$}
\label{discussion:hco+}
The strong emission of HCO$^+$ toward the region not directly associated with the shock front, B0, is consistent with previous work, where the molecule is thought to be rapidly produced in the gas phase. \citet{Rawlings2004} found that HCO$^+$ is enhanced in the boundary layer between the jet and the surrounding ambient gas, which is observed here. Within this region, the shock will cause H$_2$O and CH$_3$OH to be lifted from the grains. With these species abundant in the gas phase, sufficient amounts of the ions H$_3$O$^+$ and CH$_3$OH$_2^+$ can be formed to rapidly produce HCO$^+$. Once HCO$^+$ resides in this high-temperature, water-rich environment, it may then be efficiently destroyed through dissociative recombination \citep{Bachiller2001} or through a reaction with H$_2$O \citep{Bergin1998}. \citet{Bergin1998} found that HCO$^+$ could be efficiently converted into H$_3$O$^+$ and CO on the order of 100 years, or the approximate timescale of the shock passage modeled.
This points to the eastern wall of C2 being the site of a recent interaction, as observed by the wing feature of HCN and HCO$^+$ seen towards this region. Then, as the shock evolves, the HCO$^+$ is rapidly destroyed, as evidenced by the weaker emission toward B1 and B2. 

\subsubsection{HNCO}
\label{discussion:hnco}

HNCO has been studied in gas-grain warm-up chemical models of hot corinos, and was found to be formed through the gas-phase destruction of more complex molecules after they are lifted from the grain surface \citep{Garrod2008, Quan2010}. In the high temperature phase of the hot core models, the gas-phase abundances were found to be only 2$\times$10$^{-9}$. However in some models, the peak abundance on the grain prior to desorption due to the warm up, was as high as $\sim$10$^{-8}$--10$^{-7}$. 

Hypothetically, this would imply that, in a hot-core-like environment, the observed abundances could be explained by all of the HNCO on the grain suddenly being lifted from the grain. However, \citet{Rodriguez2010} argue that the gas-phase chemistry within shocks should differ due its higher temperatures and densities over a short period of time. They propose the dominant pathways to be a) the initial grain erosion increasing the HNCO then b) neutral-neutral gas phase reaction of CN and O$_2$ to produce OCN, which may be favored due to the significant enhancement of O$_2$ within the post-shock gas \citep{Bergin1998, Gusdorf2008}. The OCN can then be rapidly converted into HNCO. 

This formation pathway is consistent with the fact that HNCO is currently the only non-sulfur-bearing molecule observed to have significantly greater enhancement in B2 than B1. This is because the efficient formation of both HNCO and several sulfur-bearing species (i.e. SO and SO$_2$) heavily relies on molecules reacting with the abundant species O$_2$, converting CN to HNCO, and H$_2$S to SO and SO$_2$. CN and HNCO have anticorrelated physical distributions, supporting this pathway \citep{Rodriguez2010}. Further observations of other species associated with the O$_2$ chemistry in shocks may help constrain the significance of this pathway. Again, the cause of this enhancement may also be due to B2's high shock velocity compared to B1.

\section{Conclusions}
\label{conclusions}
We report high-resolution interferometric maps of emission from CH$_3$OH, HCN, HCO$^+$, and HNCO toward the southern outflow of L1157 from CARMA observations. In addition to defining three new clumps (B0k, B0$\ell$, and B2d), we utilized the non-LTE code \textsc{radex} to constrain the column densities and abundances of three of our species. We find:
\begin{itemize}
	\item The abundance of CH$_3$OH is relatively consistent across the regions and appears to be primarily produced by the liberation off of grain surfaces due to shocks
	\item Both HCN/HCO$^+$ and CH$_3$OH/HNCO display velocity profiles that are indicative of their originating from the same physical regions
	\item The east/west chemical differentiation observed in B0 and B1 could be explained by the difference in shock-chemistry that occurs when the impacted medium is cold, dense, and quiescent or warm, diffuse, and previously-shocked, as evidenced by the molecular enhancement and velocity profiles
	\item Through the first interferrometric maps of HNCO, enhanced abundances are seen toward B2 relative to B1, which may be explained by the importance of the formation of O$_2$ in shocked regions or by differences in shock velocities.
\end{itemize}

\acknowledgments
EH, AMB\added{,} and CNS thank the National Science Foundation (NSF) for continuing to support the astrochemistry program at the University of Virginia and the NASA Exobiology and Evolutionary Biology Program through a subcontract from Rensselaer Polytechnic Institute.
NMD gratefully acknowledges funding by the National Radio Astronomy Observatory Summer Research Internship, a Research Experience for Undergraduates funded by the NSF and Peter Teuben for helpful discussion.
BAM gratefully acknowledges funding by an NSF Graduate Research Fellowship during initial portions of this work.  The National Radio Astronomy Observatory is a facility of the National Science Foundation operated under cooperative agreement by Associated Universities, Inc.
Portions of this material are based upon work at the Caltech Submillimeter Observatory, which was operated by the California Institute of Technology under cooperative agreement with the National Science Foundation (AST-0838261).  Support for CARMA construction was derived from the Gordon and Betty Moore Foundation, the Kenneth T. and Eileen L. Norris Foundation, the James S. McDonnell Foundation, the Associates of the California Institute of Technology, the University of Chicago, the states of California, Illinois, and Maryland, and the National Science Foundation. CARMA development and operations were supported by the National Science Foundation under a cooperative agreement, and by the CARMA partner universities.

\appendix
\section{Spectra for All Molecules and Regions}
\label{appendix:spectra}
For each of the species observed with CARMA (see Table \ref{tab:carma_transitions}), the spectra for all of the defined regions in Table \ref{tab:clumps_regions} are provided below. For species with multiple transitions, CH$_3$OH and HCN, the velocities displayed are relative to the rest frequency of the central transition.\added{ For HCN, the unknown feature discussed in Section \ref{spectra:hcn} is marked with a blue dashed line.}

\begin{figure*}[h!]
\centering
\includegraphics[height=0.9\textheight]{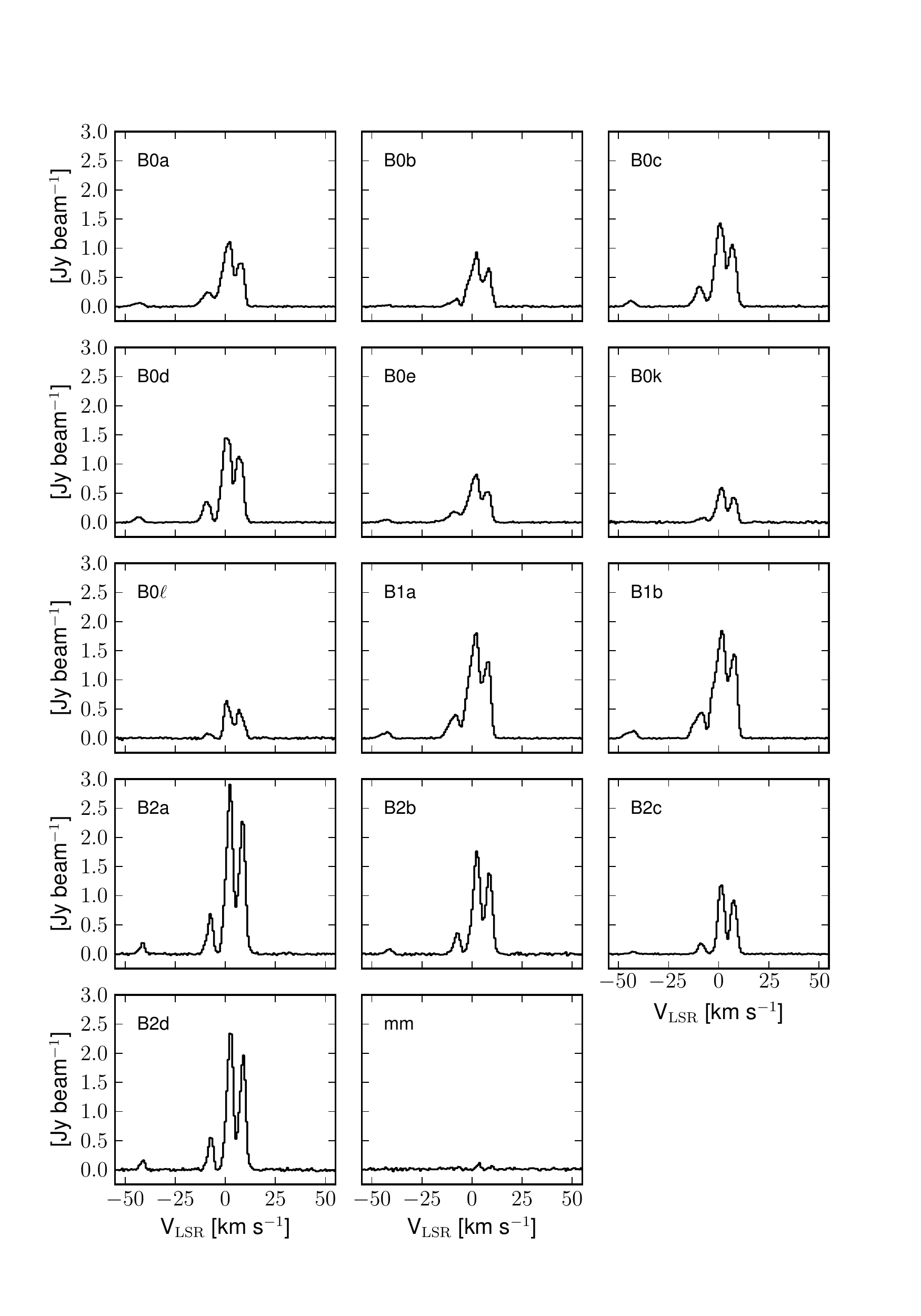} 
\caption{Extracted spectra of all regions in Table \ref{tab:clumps_regions} for CH$_3$OH emission from the CARMA observations.}
\label{fig:spectra_ch3oh}
\end{figure*}

\clearpage

\begin{figure}
\centering
\includegraphics[height=0.9\textheight]{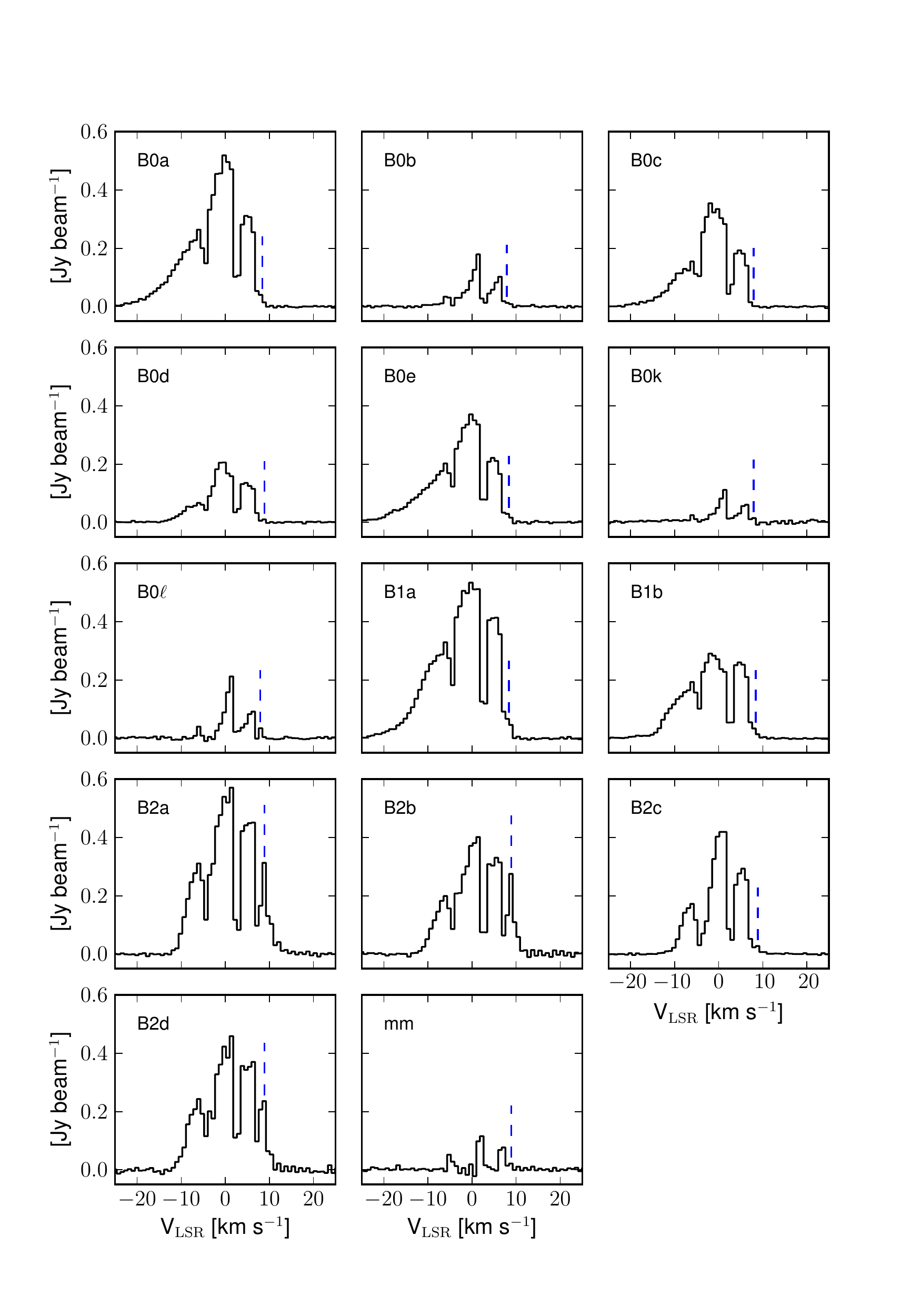} 
\caption{Extracted spectra of all regions in Table \ref{tab:clumps_regions} for HCN emission from the CARMA observations.\added{ The unknown feature is marked with a blue dashed line.}}
\label{fig:spectra_hcn}
\end{figure}

\clearpage

\begin{figure}
\centering
\includegraphics[height=0.9\textheight]{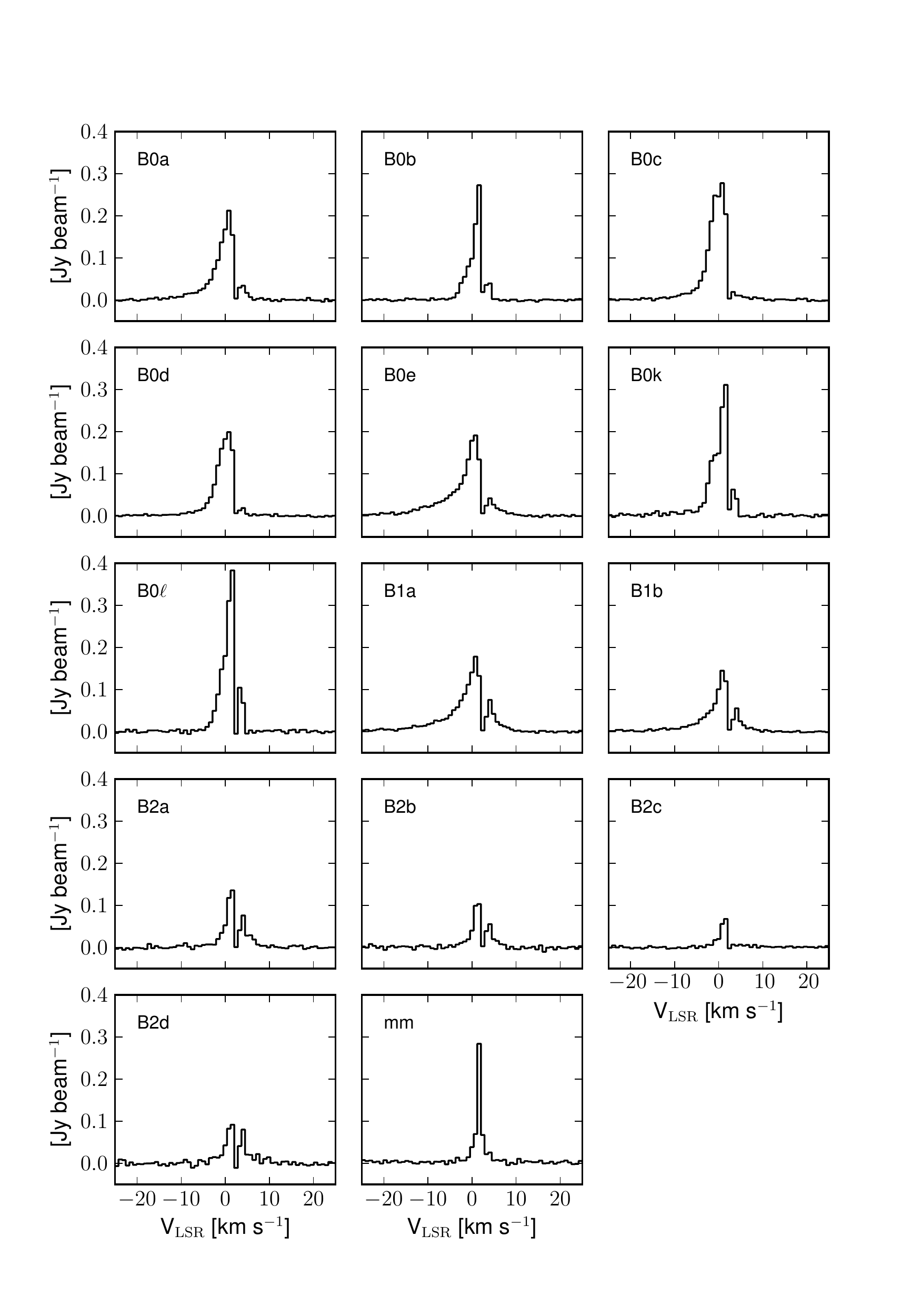} 
\caption{Extracted spectra of all regions in Table \ref{tab:clumps_regions} for HCO$^{+}$ emission from the CARMA observations.}
\label{fig:spectra_hco}
\end{figure}

\clearpage

\begin{figure}
\centering
\includegraphics[height=0.9\textheight]{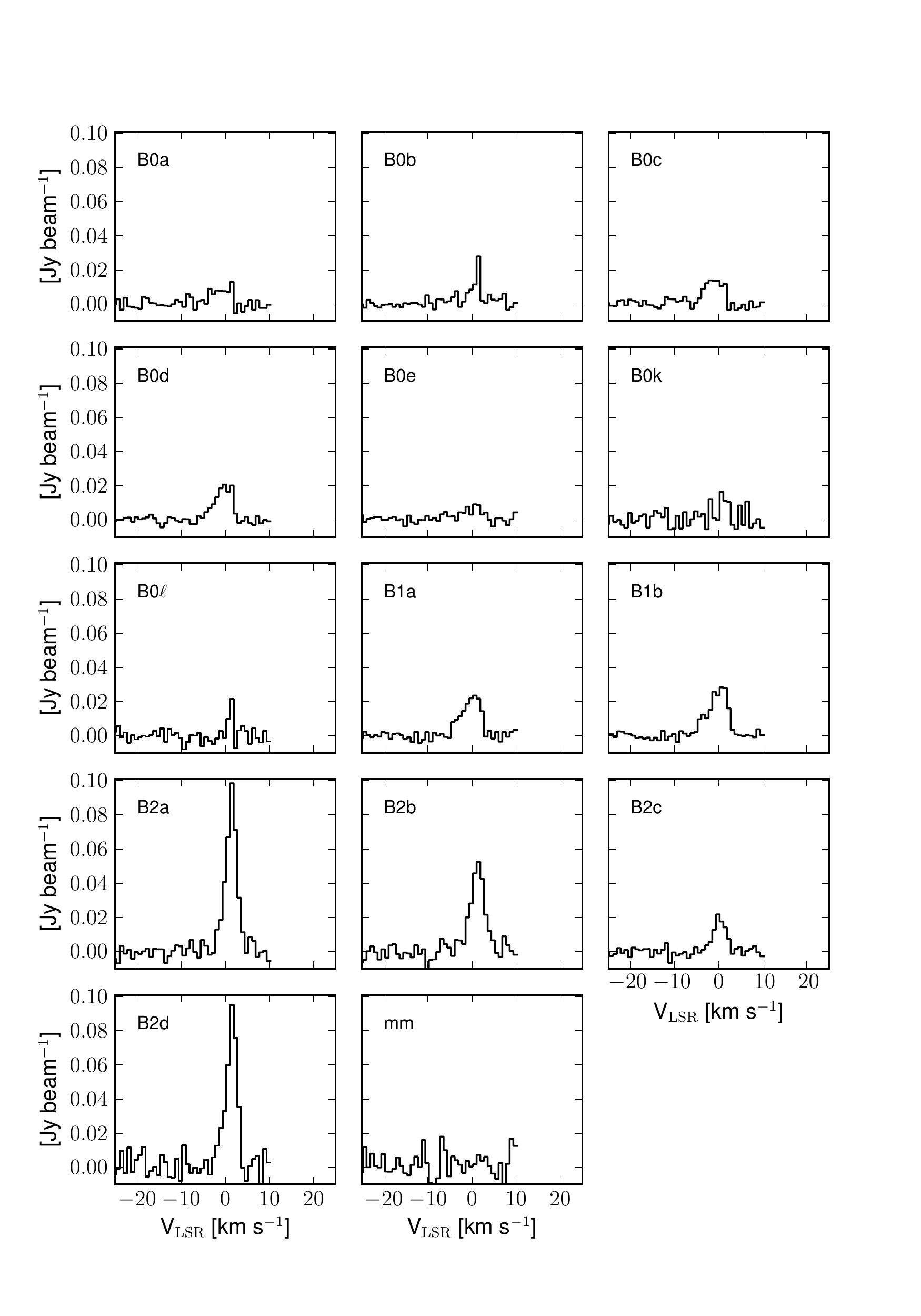} 
\caption{Extracted spectra of all regions in Table \ref{tab:clumps_regions} for HNCO emission from the CARMA observations.}
\label{fig:spectra_hnco}
\end{figure}

\clearpage

\section{Tables of Spectral Fits for All Molecules and Regions}
\label{appendix:fits}
For each of the species observed with CARMA (see Table \ref{tab:carma_transitions}), the parameters of the Gaussian fits of the spectra are tabulated below for all of the defined regions in Table \ref{tab:clumps_regions}. Gaussian profiles were generated utilizing a modified version of the line fitting code described in \citet{Corby2015}. \added{The fourth column in Tables \ref{tab:spectral_fit_ch3oh} and \ref{tab:spectral_fit_hcn} and third column in Table \ref{tab:spectral_fit_hnco} shows the LSR velocity of the line peak.} For HCN, the unknown feature seen in B2 is listed as the ``4$^{\text{th}}$ component.''

\begin{deluxetable}{l c c c c}
\tablewidth{0pt}
\tablecaption{CH$_3$OH Spectral Fits}
\tablecolumns{8}
\tablehead{ Region & Transition & Peak & \added{$v_{\text{LSR}}$} & Line Width \\
  &   & (K) & (km s$^{-1}$) & (km s$^{-1}$) }
\startdata
B0a & $2_{-1,2}-1_{-1,1}$   &  2.92  &  1.30  &  3.63 \\ 
    & $2_{ 0,2}-1_{ 0,1}$++ &  4.25  &  1.02  &  6.26 \\ 
    & $2_{ 0,2}-1_{ 0,1}$   &  0.91  &  0.81  &  7.03 \\ 
    & $2_{ 1,1}-1_{ 1,0}$   &  0.24  &  0.00  &  5.48 \\ 
B0b & $2_{-1,2}-1_{-1,1}$   &  2.37  &  1.70  &  3.32 \\ 
    & $2_{ 0,2}-1_{ 0,1}$++ &  3.20  &  1.30  &  6.82 \\ 
    & $2_{ 0,2}-1_{ 0,1}$   &  0.42  &  0.59  &  5.05 \\ 
    & $2_{ 1,1}-1_{ 1,0}$   &  0.08  &  0.28  &  5.86 \\ 
B0c & $2_{-1,2}-1_{-1,1}$   &  4.13  &  0.46  &  3.94 \\ 
    & $2_{ 0,2}-1_{ 0,1}$++ &  5.60  &  0.28  &  5.58 \\ 
    & $2_{ 0,2}-1_{ 0,1}$   &  1.31  &  -0.15  &  5.14 \\ 
    & $2_{ 1,1}-1_{ 1,0}$   &  0.38  &  -0.46  &  4.59 \\ 
B0d & $2_{-1,2}-1_{-1,1}$   &  4.58  &  0.62  &  4.12 \\ 
    & $2_{ 0,2}-1_{ 0,1}$++ &  5.93  &  0.40  &  5.33 \\ 
    & $2_{ 0,2}-1_{ 0,1}$   &  1.44  &  0.06  &  4.25 \\ 
    & $2_{ 1,1}-1_{ 1,0}$   &  0.36  &  0.06  &  4.37 \\ 
B0e & $2_{-1,2}-1_{-1,1}$   &  2.03  &  1.36  &  3.59 \\ 
    & $2_{ 0,2}-1_{ 0,1}$++ &  3.08  &  1.21  &  6.23 \\ 
    & $2_{ 0,2}-1_{ 0,1}$   &  0.66  &  1.46  &  9.48 \\ 
    & $2_{ 1,1}-1_{ 1,0}$   &  0.18  &  0.34  &  4.65 \\ 
B0k & $2_{-1,2}-1_{-1,1}$   &  1.76  &  0.96  &  3.59 \\ 
    & $2_{ 0,2}-1_{ 0,1}$++ &  2.32  &  0.87  &  4.77 \\ 
    & $2_{ 0,2}-1_{ 0,1}$   &  0.27  &  1.58  &  6.69 \\ 
    & $2_{ 1,1}-1_{ 1,0}$   &  0.05  &  0.19  &  3.81 \\ 
B0$\ell$ & $2_{-1,2}-1_{-1,1}$   &  1.77  &  0.53  &  4.80 \\ 
    & $2_{ 0,2}-1_{ 0,1}$++ &  2.51  &  0.56  &  3.59 \\ 
    & $2_{ 0,2}-1_{ 0,1}$   &  0.31  &  0.96  &  3.66 \\ 
    & $2_{ 1,1}-1_{ 1,0}$   &  -     &  -     &  -    \\ 
B1a & $2_{-1,2}-1_{-1,1}$   &  4.78  &  1.39  &  3.38 \\ 
    & $2_{ 0,2}-1_{ 0,1}$++ &  6.68  &  1.05  &  7.00 \\ 
    & $2_{ 0,2}-1_{ 0,1}$   &  1.45  &  0.40  &  6.35 \\ 
    & $2_{ 1,1}-1_{ 1,0}$   &  0.38  &  0.31  &  5.36 \\ 
B1b & $2_{-1,2}-1_{-1,1}$   &  5.16  &  1.24  &  3.41 \\ 
    & $2_{ 0,2}-1_{ 0,1}$++ &  6.79  &  0.81  &  7.44 \\ 
    & $2_{ 0,2}-1_{ 0,1}$   &  1.63  & -0.34  &  6.10 \\ 
    & $2_{ 1,1}-1_{ 1,0}$   &  0.47  & -0.28  &  5.98 \\ 
B2a & $2_{-1,2}-1_{-1,1}$   &  9.04  &  1.86  &  3.53 \\ 
    & $2_{ 0,2}-1_{ 0,1}$++ & 10.80  &  1.83  &  4.18 \\ 
    & $2_{ 0,2}-1_{ 0,1}$   &  2.57  &  1.89  &  3.41 \\ 
    & $2_{ 1,1}-1_{ 1,0}$   &  0.72  &  1.98  &  3.04 \\ 
B2b & $2_{-1,2}-1_{-1,1}$   &  5.56  &  2.01  &  3.63 \\ 
    & $2_{ 0,2}-1_{ 0,1}$++ &  6.66  &  1.98  &  4.49 \\ 
    & $2_{ 0,2}-1_{ 0,1}$   &  1.38  &  2.05  &  3.66 \\ 
    & $2_{ 1,1}-1_{ 1,0}$   &  0.32  &  1.98  &  3.66 \\ 
B2c & $2_{-1,2}-1_{-1,1}$   &  3.76  &  0.87  &  3.63 \\ 
    & $2_{ 0,2}-1_{ 0,1}$++ &  4.79  &  0.87  &  3.78 \\ 
    & $2_{ 0,2}-1_{ 0,1}$   &  0.71  &  0.74  &  3.75 \\ 
    & $2_{ 1,1}-1_{ 1,0}$   &  0.15  &  0.74  &  3.59 \\ 
B2d & $2_{-1,2}-1_{-1,1}$   &  7.67  &  2.11  &  3.38 \\ 
    & $2_{ 0,2}-1_{ 0,1}$++ &  9.21  &  2.08  &  3.87 \\ 
    & $2_{ 0,2}-1_{ 0,1}$   &  2.24  &  2.14  &  3.04 \\ 
    & $2_{ 1,1}-1_{ 1,0}$   &  0.63  &  2.14  &  2.91 \\ 
mm  & $2_{-1,2}-1_{-1,1}$   &  0.22  &  3.10  &  2.88 \\ 
    & $2_{ 0,2}-1_{ 0,1}$++ &  0.41  &  3.01  &  2.51 \\ 
    & $2_{ 0,2}-1_{ 0,1}$   &  -     &  -     &  -    \\ 
    & $2_{ 1,1}-1_{ 1,0}$   &  -     &  -     &  -    
\enddata
\label{tab:spectral_fit_ch3oh}
\end{deluxetable}
\placetable{tab:spectral_fit_ch3oh}

\begin{deluxetable}{l c c c c}
\tablewidth{0pt}
\tablecaption{HCN Spectral Fits}
\tablecolumns{8}
\tablehead{ Region & Transition & Peak & \added{$v_{\text{LSR}}$} & Line Width \\
  &   & (K) & (km s$^{-1}$) & (km s$^{-1}$) }
\startdata
B0a & $J=1-0,F=1-1$ & 4.19  &  0.81  &  3.50 \\
    & $J=1-0,F=2-1$ & 6.11  &  0.10  &  5.90 \\
    & $J=1-0,F=0-1$ & 3.05  &  0.85  &  8.00 \\
B0b & $J=1-0,F=1-1$ & 1.30  &  1.29  &  2.10 \\
    & $J=1-0,F=2-1$ & 2.74  &  1.12  &  2.30 \\
    & $J=1-0,F=0-1$ & 0.54  &  4.90  &  2.80 \\
B0c & $J=1-0,F=1-1$ & 2.42  &  0.68  &  3.50 \\
    & $J=1-0,F=2-1$ & 4.34  & -0.47  &  5.70 \\
    & $J=1-0,F=0-1$ & 1.69  & -0.27  &  7.50 \\
B0d & $J=1-0,F=1-1$ & 1.74  &  0.71  &  3.70 \\
    & $J=1-0,F=2-1$ & 2.36  & -0.20  &  5.50 \\
    & $J=1-0,F=0-1$ & 0.79  &  0.54  &  5.60 \\
B0e & $J=1-0,F=1-1$ & 2.79  &  0.74  &  3.70 \\
    & $J=1-0,F=2-1$ & 4.04  &  0.17  &  6.50 \\
    & $J=1-0,F=0-1$ & 2.49  &  2.16  &  8.00 \\
B0k & $J=1-0,F=1-1$ & 0.98  &  1.32  &  2.20 \\
    & $J=1-0,F=2-1$ & 1.62  &  1.25  &  2.20 \\
    & $J=1-0,F=0-1$ & 0.38  &  1.52  &  1.20 \\
B0$\ell$ & $J=1-0,F=1-1$ & 1.64  &  1.22  &  2.10 \\
    & $J=1-0,F=2-1$ & 2.79  &  1.32  &  2.10 \\
    & $J=1-0,F=0-1$ & 0.46  &  1.45  &  1.30 \\
B1a & $J=1-0,F=1-1$ & 6.02  &  0.81  &  3.40 \\
    & $J=1-0,F=2-1$ & 6.56  &  0.03  &  6.10 \\
    & $J=1-0,F=0-1$ & 3.14  &  0.41  &  7.80 \\
B1b & $J=1-0,F=1-1$ & 3.56  &  0.71  &  3.40 \\
    & $J=1-0,F=2-1$ & 3.54  & -0.41  &  5.90 \\
    & $J=1-0,F=0-1$ & 1.97  &  0.07  &  7.50 \\
B2a & $4^{\text{th}}$ component & 5.25  &  4.50  &  1.60 \\
    & $J=1-0,F=1-1$ & 5.95  &  0.88  &  3.50 \\
    & $J=1-0,F=2-1$ & 6.32  &  0.17  &  5.70 \\
    & $J=1-0,F=0-1$ & 3.42  &  0.88  &  4.70 \\
B2b & $4^{\text{th}}$ component & 3.57  &  4.43  &  1.80 \\
    & $J=1-0,F=1-1$ & 4.27  &  0.85  &  3.50 \\
    & $J=1-0,F=2-1$ & 4.49  &  0.54  &  5.50 \\
    & $J=1-0,F=0-1$ & 2.08  &  1.89  &  5.00 \\
B2c & $J=1-0,F=1-1$ & 3.94  &  0.85  &  3.30 \\
    & $J=1-0,F=2-1$ & 5.61  &  0.47  &  4.00 \\
    & $J=1-0,F=0-1$ & 2.08  &  0.71  &  3.90 \\
B2d & $4^{\text{th}}$ component & 3.53  &  4.23  &  1.70 \\
    & $J=1-0,F=1-1$ & 4.76  &  0.88  &  3.70 \\
    & $J=1-0,F=2-1$ & 5.48  &  0.30  &  5.00 \\
    & $J=1-0,F=0-1$ & 2.73  &  1.01  &  4.50 \\
mm  & $J=1-0,F=1-1$ & 1.17  &  2.44  &  1.52 \\
    & $J=1-0,F=2-1$ & 1.79  &  2.33  &  1.35 \\
    & $J=1-0,F=0-1$ & 0.72  &  2.54  &  1.25
\enddata
\label{tab:spectral_fit_hcn}
\end{deluxetable}
\placetable{tab:spectral_fit_hcn}

\begin{deluxetable}{l c c c}
\tablewidth{0pt}
\tablecaption{HNCO Spectral Fits}
\tablecolumns{8}
\tablehead{ Region  & Peak & \added{$v_{\text{LSR}}$} & Line Width \\
    & (K) & (km s$^{-1}$) & (km s$^{-1}$) }
\startdata
B0a &  0.13  & -0.68  &  4.88 \\ 
B0b &  0.38  &  1.74  &  1.30 \\ 
B0c &  0.21  & -0.75  &  4.47 \\ 
B0d &  0.29  &  0.03  &  4.30 \\ 
B0e &  0.10  &  0.58  &  4.91 \\ 
B0k &  0.22  &  1.60  &  1.94 \\ 
B0$\ell$ &  0.36  &  1.60  &  0.89 \\ 
B1a &  0.32  &  0.27  &  4.91 \\ 
B1b &  0.38  &  0.38  &  4.94 \\ 
B2a &  1.26  &  1.81  &  2.93 \\ 
B2b &  0.70  &  1.70  &  3.65 \\ 
B2c &  0.28  &  0.61  &  3.17 \\ 
B2d &  1.26  &  1.91  &  2.76 \\ 
mm  &  -     &  -     &  -     
\enddata
\label{tab:spectral_fit_hnco}
\end{deluxetable}
\placetable{tab:spectral_fit_hnco}

\end{document}